\documentclass{article}
\usepackage{color, xcolor, colortbl}
\usepackage{graphicx,epstopdf}
\usepackage[margin=1in]{geometry} 
\usepackage{enumerate}
\usepackage{amsmath,amssymb,amsthm}
\usepackage{caption}
\usepackage{comment}
\usepackage{subcaption}
\usepackage{bm}
\usepackage{appendix}
\usepackage{enumitem}
\usepackage{multirow}
\usepackage{braket}
\usepackage[english]{babel}
\usepackage{hyperref}
\usepackage[ruled,vlined]{algorithm2e}
\usepackage{algpseudocode}
\usepackage[capitalize]{cleveref}
\crefname{algorithm}{Algorithm}{Algorithms}
\Crefname{algorithm}{Algorithm}{Algorithms}
\usepackage[sort&compress,square,numbers]{natbib}
\usepackage{adjustbox}
\usepackage{xspace}
\usepackage[roman]{complexity}
\usepackage{soul}

\usepackage{quantikz}
\usepackage{authblk}

\SetKw{KwPro}{Process:}
\newcommand{\wdd}{\mathrm{wd}}
\newcommand{\gd}{\mathrm{gd}}

\newcommand{\ad}{\operatorname{ad}}

\newcommand{\norm}[1]{\left\lVert#1\right\rVert}

\newcommand{\Or}{\mathcal{O}}

\newtheorem{thm}{\protect\theoremname}
\theoremstyle{plain}
\newtheorem{lemma}[thm]{\protect\lemmaname}
\theoremstyle{plain}
\newtheorem{rem}[thm]{\protect\remarkname}
\theoremstyle{plain}
\theoremstyle{plain}
\newtheorem{prop}[thm]{\protect\propositionname}
\theoremstyle{plain}
\newtheorem{cor}[thm]{\protect\corollaryname}

\newtheorem{defn}[thm]{\protect\definitionname}

\providecommand{\definitionname}{Definition}
\providecommand{\assumptionname}{Assumption}
\providecommand{\corollaryname}{Corollary}
\providecommand{\lemmaname}{Lemma}
\providecommand{\propositionname}{Proposition}
\providecommand{\remarkname}{Remark}
\providecommand{\theoremname}{Theorem}

\newcommand\blfootnote[1]{%
  \begingroup
  \renewcommand\thefootnote{}\footnote{#1}%
  \addtocounter{footnote}{-1}%
  \endgroup
}

\title{High-order Magnus Expansion for Hamiltonian Simulation}

\author[1,2]{Di Fang}
\author[3]{Diyi Liu}
\author[1,2]{Shuchen Zhu}
\affil[1]{Department of Mathematics, Duke University, Durham, NC 27710, USA}
\affil[2]{Duke Quantum Center, Duke University, Durham, NC 27701, USA}
\affil[3]{ Applied Mathematics and Computational Research Division, 
Lawrence Berkeley National Laboratory, Berkeley, California 94720, USA}

\date{}

\begin{document}

\tikzset{
  U/.style={
    draw,
    rectangle,
    minimum width=0.5cm,
    minimum height=0.5cm,
    fill=white
  },
  Uf/.style={
    draw,
    rectangle,
    minimum width=0.25cm,
    minimum height=0.25cm,
    fill=black
  },
  emptyctl/.style={       
    fill=none,          %
    inner sep=0pt,      %
    minimum size=0pt    %
  }
}

\maketitle

\begin{abstract}
Efficient simulation of quantum dynamics with time-dependent Hamiltonians is important not only for time-varying systems but also for time-independent Hamiltonians in the interaction picture. Such simulations are more challenging than their time-independent counterparts due to the complexity introduced by time ordering. Existing algorithms that aim to capture commutator-based scaling either exhibit polynomial cost dependence on the Hamiltonian’s time derivatives or are limited to low-order accuracy.
In this work, we establish the general commutator-scaling error bounds for the truncated Magnus expansion at arbitrary order, where only Hamiltonian terms appear in the nested commutators, with no time derivatives involved. Building on this analysis, we design a high-order quantum algorithm with explicit circuit constructions. The algorithm achieves cost scaling with the commutator structure in the high-precision regime and depends only logarithmically on the Hamiltonian’s time variation, making it efficient for general time-dependent settings, including the interaction picture.
\end{abstract}

\setcounter{tocdepth}{2}
\tableofcontents

\blfootnote{Emails: di.fang@duke.edu; diyiliu@lbl.gov; shuchen.zhu@duke.edu.}

\section{Introduction}

Simulating quantum systems with time-dependent Hamiltonians is an important task of quantum computation. 
Many quantum processes naturally involve explicitly time-dependent Hamiltonians due to external controls or interactions with changing environments~\cite{VandersypenChuang2004,BrifChakrabartiRabitz2010,DongPetersen2010,del2013,finvzgarLukinSels2025,ZhangPagnoHessEtAl2017,JurcevicShenHaukeEtAl2017,BernienSchwartzSylvain2017}.
In addition, such systems also arise from algorithmic design. For instance, adiabatic quantum computing and quantum optimization~\cite{FarhiGoldstoneGutmannEtAl2000,AlbashLidar2018,AbbasAmbainisAugustinoEtAl2024} are inherently formulated with time-dependent Hamiltonians and have broad applications across a wide range of computational problems. Another notable example is the interaction picture~\cite{LowWiebe2019,RajputRoggeroWiebe2021}: even when the underlying Hamiltonian is time-independent, transforming the problem into a time-dependent form via the interaction picture can often lead to more efficient algorithms. 
Accurately simulating these scenarios demands efficient quantum algorithms capable of handling the general time-ordered evolution operator $U(t) = \mathcal{T}\exp\left(-i\int_0^t H(\tau)d\tau\right)$, which presents a central challenge absent in the well-studied time-independent case.

For time-dependent Hamiltonian simulation, the quality of an algorithm is measured by how its error and resource costs depend on two main factors: the time derivative (or variation) of $H$ and the spectral norm of $H$, or, when achievable, the commutators of $H$.
A key challenge is that rapidly varying $H(t)$ can make simulation costly, especially if the algorithm’s complexity scales polynomially with $H'(t)$. Ideally, only weak (such as logarithmic) dependence on $H'(t)$ should appear. 
In the analysis of simulation algorithms, error and cost bounds are typically expressed in terms of the operator norm of $H$, however, when commutator-based bounds are achievable, they often yield stronger results and are therefore more desirable.
Commutator scaling leverages the structure of $H(t)$: if terms commute or nearly commute, the relevant commutator norms are much smaller than what operator norm-based analysis would suggest, leading to tighter error bounds. 
For the time-independent case, Trotter-based algorithms achieve commutator scaling~\cite{ChildsSuTranEtAl2020, WiebeBerryHoyerEtAl2010, ChildsSu2019, SahinogluSomma2020, AnFangLin2021, SuHuangCampbell2021, ZhaoZhouShawEtAk2021, BornsWeilFang2022,GongZhouLi2023, LowSuTongTran2023, ZhaoZhouChilds2024, MoralesCostaEtAl2025, LowKliuchnikovWiebe2019}.
In the general time-dependent case, however, the standard Trotter formula~\cite{Lloyd1996} does not exhibit commutator scaling, and the generalized Trotter formula~\cite{HuyghebaertDeRaedt1990}--which decomposes the Hamiltonian into time-ordered exponentials--achieves it only at low orders~\cite{AnFangLin2021}. Recent significant advances~\cite{MizutaIkedaFujii2024} have further shown that at higher orders, the generalized Trotter formula yields error bounds that involve both commutators and the Hamiltonian's time derivatives that are not necessarily in a commutator form. In regimes where the derivative contributions are negligible, the methods can benefit from the commutator cancellations. While the generalized Trotter approach assumes that each individual time-ordered exponential can be efficiently implemented, it can offer appealing cost estimates when the Hamiltonian takes the form $H(t) = \sum_j f_j(t) H_j$, with scalar functions $f_j(t)$ that do not vary much in time.
In practice, a good algorithm ideally is expected to balance these dependencies, incorporating commutator structure, and avoiding polynomial scaling with the time derivative if possible.

\subsection{Main results}
\begin{table}[h!]\centering
 \resizebox{0.98\textwidth}{!}
 {
\begin{tabular}{ll|cccc}
 &  & \begin{tabular}{l}
      Weak dependence  \\
      on $\|\partial_t^p H(t)\|$
    \end{tabular} &   \begin{tabular}{l}
      Commutator  \\
      scaling\footnotemark
    \end{tabular}  & 
    \begin{tabular}{l}Arbitrary \\ High order\end{tabular}  & \begin{tabular}{l}
      Efficient in the  \\
      interaction picture
    \end{tabular}\\
  \cline{1-6}
Trotter-type algorithms~\cite{ChildsSuTranEtAl2020,WiebeBerryHoyerEtAl2010,MizutaIkedaFujii2024,Mizuta2025,LowKliuchnikovWiebe2019,AftabAnTrivisa2024} &   & $\times$ &    $\times$  & \textcolor{black}{\checkmark} & $\times$\\
Truncated Dyson-based~\cite{BerryChildsCleveEtAl2015,LowWiebe2019,KieferovaSchererBerry2019,BerryChildsSuEtAl2020} &  &  \textcolor{black}{\checkmark}  &   $\times$ & \textcolor{black}{\checkmark} & \textcolor{black}{\checkmark}\\
Randomized methods~\cite{BerryChildsSuEtAl2020,PoulinQarrySommaVerstraete2011} &  & \textcolor{black}{\checkmark} & $\times$ & $\times$&\textcolor{black}{\checkmark}\\
Discrete clock~\cite{WatsonWatkins2024, LiWang2025} &  &\textcolor{black}{\checkmark}    & $\times$&\textcolor{black}{\checkmark} & $\times$ \\
Classical Magnus integrators~\cite{CasaresZiniArrazola2024} &  & $\times$ &   $\times$ & \textcolor{black}{\checkmark}& $\times$\\
Prior quantum Magnus integrators~\cite{AnFangLin2022,FangLiuSarkar2025,Borns-WeilFangZhang2025}&  & \textcolor{black}{\checkmark} 
&\textcolor{black}{\checkmark} & $\times$ & \textcolor{black}{\checkmark} 
\\ This work &  &\textcolor{blue!90!black}{\checkmark}  & \textcolor{blue!90!black}{\checkmark} &\textcolor{blue!90!black}{\checkmark} & \textcolor{blue!90!black}{\checkmark} 
\end{tabular}
}
\caption{Comparison of cost scaling with state-of-the-art results for general $H(t)$. 
}
\label{tab:main_result_1}
\end{table}
\footnotetext{Commutator scaling refers to the expression where all error terms appear solely in commutator form, without any dependence on time derivatives.}
In this work, we present the first commutator error scaling result for the arbitrary $p$-th order Magnus expansion of underlying dynamics, and introduce a quantum algorithm with explicit circuit construction that, at high precision, achieves query complexity determined by nested commutators of the Hamiltonian itself, and requires only weak (logarithmic) dependence on the time derivative of the Hamiltonian for the number of elementary gates.
To the best of our knowledge, this is the first result among known methods to achieve such commutator scaling in the high precision regime for general time-dependent Hamiltonian simulation while maintaining only weak dependence on the time derivative. We compare our results to prior state-of-the-art results in the literature, as summarized in \cref{tab:main_result_1}.
The main results are summarized informally below.

\begin{thm}[Short-time Magnus error bound -- informal version of \cref{thm: local_magnus_error}] 
    Let $U_p(t)$ be the $p$-th order Magnus expansion of $U(t)$. Then 
\begin{equation}  
\|U_p(t)-U(t) \|
\leq
C\bar\alpha_{\mathrm{comm}}^{p+1} t^{p+1},
\end{equation}  
where $C>0$ is an absolute constant independent of $p$, and $t$ is a short time step with $\bar \alpha_\mathrm{comm}t < 1$. Here $\bar \alpha_\mathrm{comm}$ depends only on a finite number of nested commutators, defined by
    \begin{equation}
\bar \alpha_\mathrm{comm} := \max_{p+1 \leq q \leq p^2+2p} \alpha_{\mathrm{comm}, q}^{1/q},
\end{equation}
where $\alpha_{\mathrm{comm},q}$ denotes the maximum norm of nested commutators over the time interval $[0,t]$ with the Hamiltonian $H$ appearing $q$ times:
\begin{equation}
\alpha_{\mathrm{comm},q} := \sup_{\tau_1, \ldots, \tau_q \in [0,t]} \max_{C \in \mathcal{C}_q} \big\| C\big(H(\tau_1), \ldots, H(\tau_q)\big) \big\|,
 \end{equation}
where $\mathcal{C}_q$ denotes the set of all nested commutators of grade $q$. Here ``grade $q$" means that the expression contains exactly $q$ occurrences of $H(\cdot)$
(possibly evaluated at different times).
\end{thm}

\begin{thm}[Cost scaling for constant $p$-th order -- informal version of \cref{thm: total_resource_cost_constantorder}]
There exists a quantum algorithm (by explicit construction) that, given a sufficiently small $\epsilon>0$ and a final time $T>0$, implements the unitary evolution $U(T)$ up to precision $\epsilon$ and probability of failure $\epsilon$.  The algorithm's query complexity is $\tilde{\mathcal{O}}\left(\frac{p^2\bar{\alpha}_\mathrm{comm}^{1+1/p}\, T^{1+1/p}}{\epsilon^{1/p}}\right)$ and the elementary gate count is $\tilde{\mathcal{O}}\left(\frac{p^3\bar{\alpha}_\mathrm{comm}^{1+1/p}\, T^{1+1/p}}{\epsilon^{1/p}}\right)$ for any integer $p\geq 1$.
\end{thm}

In practice, the algorithm is often most useful at a finite order $p$. Nevertheless, as a byproduct of our result, we show that by further optimizing the choice of $p$, the Magnus series can achieve (near-optimal) $\polylog(1/\epsilon)$ scaling in the overall algorithmic complexity (see \cref{cor:optimal_query_complexity}). 
Note that this feature is particularly interesting when compared with high-order Trotterization that achieves commutator scaling in the time-independent case~\cite{ChildsSuTranEtAl2020}, where one expects $1/\epsilon^{o(1)}$ scaling rather than $\polylog(1/\epsilon)$. We further remark that polylogarithmic scaling in the precision parameter can also be achieved for general time-dependent Hamiltonians via the Dyson series~\cite{BerryChildsCleveEtAl2015,KieferovaSchererBerry2019,LowWiebe2019,BerryChildsSuEtAl2020} and an improved discrete clock~\cite{LiWang2025}.
However, when truncated at any fixed finite order, such algorithms do not achieve commutator scaling, in contrast to ours.

\medskip
\noindent \textbf{Analysis novelty.}
The technical novelty of our error analysis lies in establishing that the $p$-th order Magnus expansion, for any finite $p$, exhibits commutator scaling. This result is of independent interest beyond algorithmic applications, given the widespread use of the Magnus expansion as a theoretical tool in physics and chemistry.
Our proof strategy differs fundamentally from prior analyses of the Magnus series. Instead of reasoning via the series representation and bounding the remainder after truncation--which necessarily introduces infinitely many terms and thus yields error bounds depending on infinitely many commutators--we employ an exact error representation, a technique widely used in numerical analysis, which was extended to quantum algorithm analysis in~\cite{ChildsSuTranEtAl2020} and subsequently applied in, e.g., \cite{SuHuangCampbell2021,AnFangLin2021,AnFangLin2022,FangLiuSarkar2025,BornsWeilFang2022,Borns-WeilFangZhang2025}. This approach allows us to show that the error depends only on finitely many commutators.
The proof proceeds by first demonstrating that all terms up to grade $p$ cancel, and then collecting all higher-order contributions. A key step in the cancellation relies on a clever use of the properties of Bernoulli numbers. Compared with the low-order case, the high-order analysis is substantially more challenging, since it requires controlling both $n$-th order Magnus term and its time derivative for all $n \leq p$, which depend recursively on all lower-order terms in a nontrivial way; in contrast, low-order cases involve tracking only one or two terms. 

\medskip

\noindent \textbf{Circuit novelty.}
The technical novelty of our algorithm and circuit is using $\poly(p)$ cost to construct the quantum circuit for $p$-th order Magnus expansion, which potentially could have factorial complexity. 
Three key components—the state preparation for coefficients in the Magnus expansion, the state preparation for the time-ordered integral, and the construction of the SELECT oracle—can each potentially contribute an additional factorial factor.  
Indeed, if we adopt the circuit construction used in \cite{FangLiuSarkar2025} and \cite{AnFangLin2021}, both the two-qubit gate counts and circuit depth is of $\mathcal{O}(p\cdot p!)$

A necessary condition to have $\poly(p)$ simulation cost is that the circuit depth for the $p$-th order Magnus expansion is of $\poly(p)$. 
We note that any Linear Combination of Unitaries (LCU) circuit construction approach based on a SELECT oracle built with multi-control gates for each term of the Magnus expansion requires a circuit depth of $\mathcal{O}(p\cdot p!)$. 
To resolve the problem, we apply a quantum lookup table~\cite{low2024trading,ZhuSundaramLow2024} in the construction of the SELECT oracle~\cite{Wan2021exponentially,liu2025block} for Magnus expansion.  
Another challenge is preparing $\mathcal{O}(p\cdot p!)$ classical data naively may require $\mathcal{O}(p \cdot p!)$ two-qubit gates. 
However, we have an explicit formula for each of the $\mathcal{O}(p\cdot p!)$ coefficients of the $p$-th order Magnus expansion, which allows efficient classical arithmetic in $\poly(p)$ cost. 
Using the arithmetic circuit and direct sampling method introduced in \cite{liu2025block}, the coefficients of the Magnus expansion can be prepared with almost optimal subnormalization factor (close to $\sqrt{1/p!}$).
The superposition of permutation states is then constructed using the method introduced by Berry et al~\cite{berry2018improved}. 
In this way, the $\Or(p!)$ coefficients of the $p$-th order Magnus expansion terms can be prepared with only $\poly(p)$ two-qubit gates. 
However, a further challenge is that while the two-qubit cost scales polynomially in $p$, the subnormalization factor for $p$-th order terms may still grow as $\Or(p!)$. 
To address this, we developed a new state preparation method inspired by the idea in Berry et al~\cite{berry2018improved}, which constructs the time-ordered integral with a subnormalization factor of $\mathcal{O}(1/p!)$. 
This approach effectively cancels the factorial growth in the subnormalization factor for the block encoding of Magnus expansion, thereby reducing the overall simulation cost from $\Or(p!)$ to $\poly(p)$.

\subsection{State of the art in the literature}

One notable method is \textit{Trotterization}, which decomposes the unitary evolution operator into sequential exponentials of simpler terms.
Trotter-based methods achieve commutator scaling for time-independent Hamiltonians~\cite{ChildsSuTranEtAl2020,ChildsSu2019} and in several other settings studied through case-specific analyses~\cite{SahinogluSomma2020,AnFangLin2021,SuHuangCampbell2021,ZhaoZhouShawEtAk2021,ChildsLengEtAl2022,FangTres2023,BornsWeilFang2022,HuangTongFangSu2023,ZengSunJiangZhao2022,GongZhouLi2023,LowSuTongTran2023,ZhaoZhouChilds2024,ChenXuZhaoYuan2024,FangQu2025,FangWuSoffer2025}. These standard Trotter formulas can also be naturally extended to time-dependent Hamiltonians; for instance, the first-order case takes the form
\begin{equation}
    \mathcal{T} e^{-i \int_0^t H_1(s) + H_2(s) \, ds} \approx  e^{-i H_2(t_\ast) t} e^{-i H_1(t_\ast) t},
\end{equation}
for some $t_\ast \in [0,t]$. However, it has been shown that standard Trotter formulas of arbitrary order introduce a polynomial cost dependence on the time derivatives of the Hamiltonian~\cite{WiebeBerryHoyerEtAl2010}.
An alternative is given by the generalized Trotter formulas~\cite{HuyghebaertDeRaedt1990}, whose first-order form is, for example, 
\begin{equation}
    \mathcal{T} e^{-i \int_0^t H_1(s) + H_2(s) \, ds} \approx  \mathcal{T} e^{-i \int_0^t H_2(s) \, ds}   \mathcal{T} e^{-i \int_0^t H_1(s) \, ds} .
\end{equation}
Such generalized Trotter formulas achieve commutator scaling at low orders~\cite{HuyghebaertDeRaedt1990,AnFangLin2021,RajputRoggeroWiebe2021}. At higher orders, however, obtaining genuine commutator scaling in terms of Hamiltonians themselves remains challenging: dependence on time derivatives of the Hamiltonian cannot be avoided in the general case~\cite{MizutaIkedaFujii2024}. Moreover, note that the generalized Trotter formula approximates the evolution as a product of simpler time-ordered exponentials, without further fully implementing the time-ordering operator within each. To make these operators efficiently implementable without destroying commutator scaling typically requires additional structural assumptions on the Hamiltonian. Otherwise, a naive implementation of each sub-time-ordered exponential--such as picking some time in the considered interval as in the standard Trotter formulas--can again lead to a loss of commutator scaling and introduce derivative dependence in forms not captured by commutators as demonstrated in~\cite{AnFangLin2021}. Another Trotter type algorithms can achieve commutator scaling for time-independent Hamiltonian is the multiproduct formula \cite{LowKliuchnikovWiebe2019}. Interestingly, \cite{AftabAnTrivisa2024} provides an explicit commutator error bound of MPF for the time-independent case, which depends on infinitely many commutators. This was later improved in \cite{Mizuta2025}, where the dependence was reduced to finitely many commutators, with the number determined by the truncation order of the approximation (analogous to the order $p$ in our setting). In this sense, our error analysis--also involving only finitely many nested commutators--is more closely aligned with the result of \cite{Mizuta2025}, but differs in that our algorithm is based on the Magnus expansion and applies to general time-dependent Hamiltonians.

\textit{Truncated Dyson-series} algorithms approximate the time-evolution operator using a truncated Dyson series (time-ordered power series expansion) and implement it via linear combinations of unitaries.
Kieferova et al.~\cite{KieferovaSchererBerry2019} first extends the truncated Taylor-seires method~\cite{BerryChildsCleveEtAl2015} with truncated Dyson series, and shows that both query complexity and gate count scale logarithmically with the inverse error and linearly on the maximum operator norm of $H(t)$.
Berry et al.~\cite{BerryChildsSuEtAl2020} improves this by introducing a rescaled Dyson-series algorithm whose complexity depends on the $L_1$ norm instead of the maximum operator norm, potentially providing large cost savings when $H(t)$ varies over time. 

A novel \textit{discrete-clock construction} is developed in~\cite{WatkinsWiebeRoggeroLee2022} that can be combined with various simulation methods to efficiently handle time-dependence. More recently, \cite{LiWang2025} extends the method to high order, achieving logarithmic dependence on the Hamiltonian’s derivatives and attaining $\polylog(1/\epsilon)$ scaling, consistent with both the Dyson series approach and our high-order Magnus–based approaches proposed here.

The \textit{Magnus expansion} rewrites the time-evolution operator as $U(t)=\exp(\Omega(t))$, where $\Omega(t)$ is an infinite series of time integrals and nested commutators.
In~\cite{AnFangLin2022}, the qHOP algorithm was introduced, derived by dropping the time-ordering operator, which agrees with truncating the Magnus expansion at first order. 
In each small time step, $H(t)$ is replaced with an effective time-independent Hamiltonian and implemented with a single unitary. 
The resulting first-order Magnus method has an error determined by commutators of $H(t)$ for general time-dependent Hamiltonians.
Recent work~\cite{FangLiuSarkar2025,Borns-WeilFangZhang2025} extended this method to a second-order Magnus algorithm, achieving commutator error scaling and logarithmic cost dependence on $H'(t)$. The second-order improvement is significant, as it contrasts with randomized algorithms~\cite{BerryChildsSuEtAl2020,PoulinQarrySommaVerstraete2011}, which, in the time-dependent case, typically face difficulties in going beyond first order. 
This line of work shows that low-order Magnus truncation, combined with efficient block encoding techniques, enables simulation of general $H(t)$ with commutator error scaling. This can be viewed as generalizing the Trotter-type methods which exhibit commutator scaling for time-independent Hamiltonians. 
In our work, we complete this line of research by establishing the first commutator scaling result without strong derivative dependence for the Magnus expansion at arbitrary order.
It is important to note that extending such analysis beyond low order is nontrivial. On the error analysis side, it is challenging to quantify the exact commutator form while keeping track of the $p$ dependence; on the circuit side, it is necessary to keep track of the explicit dependence on the order parameter $p$ and to construct a circuit whose cost scales only mildly in $p$, which is polynomial dependence in our construction. At first glance, it may seem infeasible, since the Magnus series involves sums over permutation groups containing $p!$ terms. Indeed, insights from Trotter analysis (which has an exponential in $p$ prefactor) indicate that the factorial growth arising from the permutation group of size $p$ makes it difficult, if not impossible, to obtain a commutator bound at higher orders without introducing a large dependence (exponential or factorial) on $p$.

Several recent novel quantum algorithms have utilized the Magnus expansion together with other simulation techniques, leading to scaling advantages in a variety of applications~\cite{CasaresZiniArrazola2024,SharmaTran2024,BosseChildsEtAl2024}. The Magnus expansion is also a key tool in the design of classical geometric, structure-preserving algorithms, most notably through the development of Magnus integrators (see, e.g., the reviews~\cite{BlanesCasasOteoRos2009,HairerHochbruckIserlesLubich2006} and the book~\cite{BlanesCasas2017book}).

In \cref{tab:main_result_1}, we highlight the scaling behavior of our result compared to other methods. 
Our algorithm is designed for general time-dependent Hamiltonians, with cost scaling only logarithmically in the time derivative. This weak dependence ensures efficiency in the interaction picture~\cite{LowWiebe2019}, which allows the algorithm to be adapted to the time-independent case as well.

\subsection{Paper roadmap}

We give an overview of the high-order Magnus algorithm in \cref{sec: prelim}.
In \cref{sec: magnus_error}, we analyze the truncation error of the Magnus expansion at arbitrary order. Next, we study the quadrature (discretization) error associated with the Magnus expansion in \cref{sec: dis_error}. 
In \cref{sec: circuit}, we present quantum circuits for implementing the truncated Magnus expansion at arbitrary order and discuss their corresponding input models.
Equipped with this comprehensive analysis, we quantify the overall resource cost of our circuits in \cref{sec: cost}. 
We conclude the paper in \cref{sec: conclusion}.

\section{High-order Magnus algorithm}\label{sec: prelim}

In this section, we first review the Magnus series, and then present an overview of the high-order Magnus algorithm in \cref{alg: magnus_alg}.

We review the Magnus expansion framework for representing linear differential equations with time-dependent coefficients. 
In particular, setting $A(t) := -i H(t)$ yields the time-dependent Schr\"odinger equation. 
The differential equation
\begin{equation}
    \dot U(t) = A(t) U(t), \quad U(0) = U_0,
\end{equation}
has the formal solution given by the time-ordered exponential,
\begin{equation}
    U(t) = \mathcal{T}\exp\left(\int_0^t A(s)\, ds\right)U_0.
\end{equation}
Throughout this work, we adopt this notation formally, understanding it as representing a well-defined underlying unitary operator.
Alternatively, the Magnus expansion expresses the propagator without explicit time ordering:
\begin{equation}
  U(t)=\exp(\Omega(t))U_0,
\end{equation}
where the Magnus operator $\Omega(t)$ satisfies the differential equation
\begin{equation}\label{eq:d_Omega_infty}
  \frac{d \Omega}{dt} = \sum_{n=0}^\infty \frac{B_n}{n!} \, \ad^n_\Omega(A).
\end{equation}
Here, $B_n$ are Bernoulli numbers, with $B_1 = -1/2$. Equivalently, the operator $\Omega(t)$ admits an infinite series representation:
\begin{equation}
 \Omega(t) =\sum_{n=1}^\infty \Omega_n(t),
\end{equation} 
where 
\begin{equation}
  \Omega_1(t) =   \int_{0}^{t} A(s)\, ds,
\end{equation}
\begin{equation}   \label{eq:mag_n_omega}
  \Omega_n(t) =  \sum_{j=1}^{n-1} \frac{B_j}{j!} \,
    \sum_{
            k_1 + \cdots + k_j = n-1 \atop
            k_1 \ge 1, \ldots, k_j \ge 1}
            \, \int_0^t \,
       \ad_{\Omega_{k_j}(s)} \, \cdots \, \ad_{\Omega_{k_2}(s)} 
          \, \ad_{\Omega_{k_1}(s)} A(s) \, ds,    \qquad n \ge 2,
\end{equation}
and $B_j$ are Bernoulli numbers. While our work considers a finite-order expansion of the series, it is helpful to note that this series converges when $\int_0^t\|A(s)\|ds<\pi$ for any bounded normal operator $A(t)$. For a detailed review of the Magnus series, we refer readers to Blanes et al.~\cite{BlanesCasasOteoRos2009}.
\begin{align}
    \Omega(t) &= \int_{0}^{t} A(t_1)\, dt_1 + \frac{1}{2} \int_{0}^{t} dt_1 \int_{0}^{t_1} dt_2 \,[A(t_1),A(t_2)] \notag \\
    &+ \frac{1}{6} \int_{0}^{t} dt_1 \int_{0}^{t_1} dt_2 \int_{0}^{t_2} dt_3 
\left( [A(t_1),[A(t_2),A(t_3)]] + [A(t_3),[A(t_2),A(t_1)]] \right) +\cdots.
\end{align}

In our work, we propose a quantum algorithm based on high-order Magnus expansion (which can achieve an arbitrary high order in accuracy). In particular, for a final computational time of interest $T$, we first partition the interval into $L$ uniform subintervals $[t_j, t_{j+1}]$, where $t_j = jh$ and $h = T/L$ denotes the time step. On each short interval $[t_j, t_{j+1}]$, we approximate the exact unitary evolution, given by the time-ordered exponential, using a $p$-th order Magnus expansion (which is a matrix exponential without time ordering):
\begin{equation}\label{eq:U_exact_apprx_U_p}
        U(t_j+h,t_j) 
    = \mathcal{T}\exp\!\left(-i \int_{t_j}^{t_j+h} H(s)\, ds\right) 
    \;\;\approx\;\; \exp\!\left(\Omega_{(p)}(t_j+h, t_j)\right) = :  U_p(t_{j+1}, t_{j}),
\end{equation}
where $\Omega_{(p)}(t_{j+1}, t_j)$ is the $p$-th order short-time Magnus expansion over the time interval $[t_j, t_{j+1}]$ given by
\begin{equation}\label{eq:mag_p_unitary_tj_tj+1}
 \Omega_{(p)}(t_{j+1}, t_j) = \sum_{n=1}^p \Omega_n (t_{j+1}, t_j),
\end{equation}
where $\Omega_n(t_{j+1}, t_j)$ is defined as in \cref{eq:mag_n_omega}, except with the time integral taken over the interval $[t_j, t_{j+1}]$ instead of $[0,t]$. 
Note that in our case $A(t) = -iH(t)$ is anti-Hermitian, the $p$-th order Magnus coefficient operator $\Omega_{(p)}(t)$ is also anti-Hermitian and $i\Omega_{(p)}(t)$ Hermitian, consistent with the fact that $U_{p}= e^{-i(i\Omega_{(p)}(t))}$.
We analyze the Magnus expansion error relative to the exact dynamics by deriving an exact error representation
in \cref{sec: magnus_error}, which allows us to reveal the commutator scaling in this expansion.

In the circuit implementation, we realize a quadrature-based approximation of $\Omega_p(t_j+h, t_j)$, denoted $\tilde{\Omega}_p(t_j+h, t_j)$, where the time integrals are evaluated via quadrature (specifically, Riemann sums; see \cref{sec: dis_error} for the quadrature error analysis and \cref{sec: circuit} for the detailed circuit construction). The overall procedure is summarized in the pseudocode below. \cref{alg: magnus_alg} outlines the main steps of the high-order Magnus expansion algorithm for simulating the time evolution of a general time-dependent Hamiltonian.

\begin{algorithm}[H]
\caption{Time-dependent Hamiltonian simulation using Magnus expansion (\cref{sec: quantum_circuit_magnus}).}
\label{alg: magnus_alg}
 \KwIn{ 
 initial state $\ket{\psi}$, HAM-T oracles as block encoding for time-dependent Hamiltonian $H(t)$, total evolution time $T$, order of Magnus expansion $p$, and required precision $\epsilon$.
 }
 \KwOut{ $\epsilon$-approximation of $\mathcal{T
 }e^{-i\int_{0}^{T} H(s) ds}\ket{\psi}$.  }

\textbf{Parameter setup:} based on the input, compute the number of time steps $L$, the number of quadrature points $M$, and the total number of ancilla qubits $\mu$. The time interval $[0,T]$ is divided into $L$ time periods, separated by time grid points $0=t_0<t_1<\dots < t_{L}=T$ with $t_\ell=h\ell, \ell\in\{0,1,\cdots,L\}$ and $h=T/L$.

\textbf{Initialization:} start with the initial state $\ket{0^{\mu}} \ket{\psi}$.

\For{each time step from $t_{j-1}$ to $t_j$}{
1. For each $k\in\{1,2,\cdots,p\}$, use a generalization of the circuit in \cref{fig:omega_3_circuit} to construct a $(2\alpha^kh^k , kn_m+n_b+k\lceil \log(p)\rceil +kn_a+k,\frac{\epsilon}{Lp})$ block encoding of $\tilde{\Omega}_{k}(t_{j},t_{j-1})$, the quadrature version of ${\Omega}_{k}(t_{j},t_{j-1})$. 
Furthermore, with a LCU circuit (\cref{fig:omega_(p)_LCU}), we then have a  $(2C_{(p)}^{\gamma}\alpha h , pn_m+n_b+p\lceil \log(p)\rceil +pn_a+p,\frac{\epsilon}{L})$ block encoding of $\tilde{\Omega}_{(p)}(t_{j},t_{j-1})=\sum_{k=1}^p \tilde{\Omega}_{k}(t_{j},t_{j-1})$. \tcp{Construction of Magnus expansion}

2. Apply QSVT and OAA~\cite{GilyenSuLowEtAl2019} to obtain a $(1,pn_m+n_b+p\lceil \log(p)\rceil +pn_a+p+2, \frac{\epsilon}{L})$ block encoding of $e^{\tilde{\Omega}_{(p)}(t_{j},t_{j-1})}$. \tcp{Short-time matrix exponential }
}

Evolve the initial state $\ket{0^{\mu}} \ket{\psi}$ with $L$ short time evolution operators and obtain
\begin{equation*}
       e^{\tilde{\Omega}_{(p)}(t_{L},t_{L-1})}\cdots e^{\tilde{\Omega}_{(p)}(t_{2},t_1)}e^{\tilde{\Omega}_{(p)}(t_{1},t_0)}\ket{\psi}
\end{equation*}
as $\epsilon$-approximation of $\mathcal{T
 }e^{-i\int_0^T H(s) ds}\ket{\psi}$.
\end{algorithm}

\section{Error analysis of $p$-th order Magnus expansion}\label{sec: magnus_error}
In this section, we provide a detailed error analysis of the $p$-th order Magnus expansion. 
In \cref{sec:error_rep_com}, we first analyze the structure of the local truncation error, and show that all terms in the error with grade less than or equal to $p$ cancel by induction, so the error depends only on nested commutators with grade at least $p+1$.
Then in \cref{sec:magnus_error_bounds}, we obtain explicit upper bounds for these higher-grade terms, showing that both local and global errors are controlled by a finite set of nested commutators when $p$ is fixed, and derive error scaling in terms of the commutator norms.

\subsection{Notation}
Depending on context, we denote by $\norm{\cdot}$ either the $l^2$ norm of a wavefunction or the operator norm (spectral norm) acting on $l^2$. Norms differing from this convention are explicitly stated.
We use the adjoint operator $\ad$ defined as 
\begin{equation}
    \ad_{B}(A) := [A, B], \quad \ad_B^{k+1}(A) := \ad_B(\ad_B^{k}(A)), \quad k \geq 1.
\end{equation}

Consider an expression $L$ involving $n$-fold time integrals whose integrand is constructed from nested commutators of the operator $A(t) = -i H(t)$. 
The \textit{width} $\mathrm{wd}(L)$ of $L$ is the number of integral layers it contains, and its \textit{grade} $\mathrm{gd}(L)$ is the total number of $A$ operators appearing in its nested commutator structure. 
Specifically, a term of grade $n$ includes $n - 1$ nested commutators. 
It follows that for non-commuting operators $A$ and $B$, each expressed as time integrals involving commutators of the Hamiltonian $H(t)$ or $A(t) = -iH(t)$, one has
\begin{equation}
    \wdd(\ad_A(B)) = \wdd(A) + \wdd(B), 
    \quad 
    \gd(\ad_A(B)) = \gd(A) + \gd(B).
\end{equation}
Here, the grade $\gd$ is a standard notion, while the width $\wdd$ is a definition introduced in this work to make the analysis and presentation clearer.
We note the following useful identities regarding width and grade:
\begin{equation}
    \gd(\Omega_n) = n, \quad   \gd(\dot \Omega_n) = n, \quad \gd(S_n^{(j)}) = n,
\end{equation}
\begin{equation}
  \wdd(\Omega_n) = n , \quad \wdd(\dot \Omega_n) = n-1, \quad \wdd(S_n^{(j)}) = n-1.
\end{equation}

We denote the exact unitary dynamics as 
\begin{equation}\label{eq:def_U(t)}
    U(t) = \mathcal{T}e^{\int_0^t A(s)\, ds} = \mathcal{T}e^{-i \int_0^t H(s)\, ds},
\end{equation}
and we denote the $p$-th order short-time Magnus expansion as
\begin{equation}\label{eq:mag_p_unitary_t_only}
    U_p(t) = e^{ \Omega_{(p)}(t)}, \quad \Omega_{(p)}(t) = \sum_{n=1}^p \Omega_n (t),
\end{equation}
where $\Omega_n (t)$ is defined in \cref{eq:mag_n_omega}. 
Equivalently, $ \Omega_n(t) $ can be written as
\begin{equation}
   \Omega_1(t) = \int_0^t A(s)\, ds 
\end{equation}
\begin{equation}\label{eq:omega_n_def_in_S}
    \Omega_n(t) = \sum_{j=1}^{n-1} \frac{B_j}{j!} \int_0^t S_n^{(j)}(\tau) \, d\tau, \quad n \geq 2,
\end{equation}
where $S_n^{(j)}$ is defined as
\begin{equation}
    S_n^{(j)} = \sum_{
            k_1 + \cdots + k_j = n-1 \atop
            k_1 \ge 1, \ldots, k_j \ge 1}
            \, \,
       \ad_{\Omega_{k_j}} \, \cdots \, \ad_{\Omega_{k_2}} 
          \, \ad_{\Omega_{k_1}} (A) .
\end{equation}
Here, we drop the explicit time dependence in the formula for notational simplicity. This representation involving $S_n^{(j)}$ becomes handy in the error analysis discussion. It is worth pointing out that considering the $p$-th order Magnus expansion is not the same as truncating $\dot \Omega$ (given by~\cref{eq:d_Omega_infty}) with $p$-terms. One can verify this directly by taking the time derivative of \cref{eq:omega_n_def_in_S}, which shows the two expressions differ. This distinction introduces significant additional complexity into the calculation below, compared with the simpler case where they would coincide.

\subsection{Error representation and commutator cancellation}\label{sec:error_rep_com}
We start by considering the differential equation followed by $U_p$:
\begin{align}\label{eq:U_p_DE}
\dot U_p(t) = & \frac{d}{dt} \exp(\Omega_{(p)}(t))  = d \exp_{\Omega_{(p)}(t)}(\dot \Omega_{(p)}(t)) \, \exp(\Omega_{(p)}(t))
\\
=& d \exp_{\Omega_{(p)}(t)}(\dot \Omega_{(p)}(t)) U_p = :\tilde{A}_p(t) U_p(t).
\end{align}
For notational simplicity, we omit the explicit time dependence in the expressions that follow. The expression for $\tilde{A}_p$ is given by the series expansion
\begin{align}\label{eq:dexp_omega_C}
 \tilde{A}_p = & d \exp_{\Omega_{(p)}}(\dot \Omega_{(p)})  =    \sum_{k=0}^{\infty} \frac{1}{(k+1)!} \,
   \mathrm{ad}_{\Omega_{(p)}}^k(\dot \Omega_{(p)})
   \\
  =  & \sum_{k=0}^{p} \frac{1}{(k+1)!} \,
   \mathrm{ad}_{\Omega_{(p)}}^k(\dot \Omega_{(p)}) +g_p(\ad_{\Omega_{(p)}})(\ad_{\Omega_{(p)}}^{p+1}(\dot \Omega_{(p)}) ),
\end{align}
where the last term is the remainder term in the series expansion of the function $\frac{e^{z}- 1}{z}$
\begin{equation}\label{eq:def_g_p}
    \frac{e^{z}- 1}{z} = \sum_{k=0}^{p} \frac{1}{(k+1)!} z^k +  g_p(z) z^{p+1},
\end{equation}

Note that the exact dynamics $U(t)$ follow the differential equation
\begin{equation}\label{eq:U_DE}
   \dot U(t) = A U. 
\end{equation}
Taking the difference between \cref{eq:U_DE} and \cref{eq:U_p_DE} and use the variation of constants formula, we have
\begin{equation}\label{eq:U-Up_integral_in_tildeA-A}
    (U - U_p) (t) = \int_{0}^t \,ds \, \mathcal{T} e^{\int_{s}^t A(\tau)\, d\tau} (A(s) - \tilde{A}_p (s)) U_p(s).
\end{equation}
Therefore, it suffices to estimate 
\begin{equation}\label{eq:tildeA_p-A_def}
    \tilde{A}_p - A = 
   \underbrace{ \sum_{k=0}^{p} \frac{1}{(k+1)!} \,
   \mathrm{ad}_{\Omega_{(p)}}^k(\dot \Omega_{(p)}) - A }_{: = \Theta_p }
   + 
   g_p(\ad_{\Omega_{(p)}})(\ad_{\Omega_{(p)}}^{p+1}(\dot \Omega_{(p)}) ). 
\end{equation} Our goal is to show that this difference contains only terms with grade at least $p+1$ and width at least $p$. This structure implies that the local truncation error of the $p$-th order Magnus expansion is of the form $\alpha_p t^{p+1}$, where $\alpha_p$ denotes the norm of the $p$-layer nested commutators. The factor of $t^{p+1}$ arises because the difference $\tilde{A}_p - A$ has width at least $p$, contributing a factor of $t^p$, and the difference $U - U_p$ introduces an additional layer of time integration, yielding an extra factor of $t$. Consequently, the global error scales with $t^p$, achieving $p$-th order accuracy with nested commutator scaling as desired.

It is straightforward to verify that all terms in the remainder
\begin{equation}
 g_p(\ad_{\Omega_{(p)}})(\ad_{\Omega_{(p)}}^{p+1}(\dot \Omega_{(p)}) )
\end{equation}
have grade at least $p+2$ and width at least $p+1$. Therefore, it suffices to show that all terms in $\Theta_p$ have grade at least $p+1$ and width at least $p$.

Note that the grade of each term in $\Theta_p$ is equal to its width plus one. Hence, we only need to focus on the grade. In other words, it suffices to show that all terms in $\Theta_p$ with grade less than or equal to $p$ cancel out, which we will show by induction.
To formalize the cancellation of lower-grade terms, we establish the following proposition.

\begin{prop} \label{prop:inductive_statement}
    Let $p \in \mathbb{N}_+$. For each $1 \leq k \leq p$, the sum of all terms in $\Theta_p$ with grade $k$ vanishes.
\end{prop} 

\begin{proof}
We begin with the base case $p = 1$. Observe that
\begin{equation}
    \Theta_1 =\dot  \Omega_1 + \frac{1}{2} \ad_{\Omega_1}(\dot \Omega_1) - A
    = \ad_{\Omega_1}(A),
\end{equation}
where we have used the identity $\dot\Omega_1 = A$. Since $\ad_{\Omega_1}(A)$ has grade 2, $\Theta_1$ contains no terms of grade $k \leq 1$, and the base case holds.

Assume that for some $p \geq 1$, the inductive statement in \cref{prop:inductive_statement} holds. We aim to show that it also holds for $p+1$, i.e., all terms in $\Theta_{p+1}$ with grade less than or equal to $p+1$ cancel out.

We begin by writing
\begin{align}\label{eq:Theta_p+1}
    \Theta_{p+1} = &
    \sum_{k=0}^{p} \frac{1}{(k+1)!} \,
   \mathrm{ad}_{\Omega_{(p+1)}}^k(\dot \Omega_{(p+1)}) - A +  \frac{1}{(p+2)!} \,
   \mathrm{ad}_{\Omega_{(p+1)}}^{p+1}(\dot \Omega_{(p+1)}).
\end{align}
The last term has grade at least $p+2$, so we only need to analyze the contributions from the first two terms.
Using the decomposition $\Omega_{(p+1)} = \Omega_{(p)} + \Omega_{p+1}$, we can expand $\ad_{\Omega{(p+1)}}^k(\dot \Omega_{(p+1)})$ in \cref{eq:Theta_p+1} as follows:
\begin{align}
  & \sum_{k=0}^{p} \frac{1}{(k+1)!} \,   \mathrm{ad}_{\Omega_{(p+1)}}^k(\dot \Omega_{(p+1)}) - A
  \\
  =& 
   \sum_{k=0}^{p} \frac{1}{(k+1)!} \, \mathrm{ad}_{\Omega_{(p)} 
   + \Omega_{p+1}}^k(\dot \Omega_{(p)}) 
   -A 
   +  \sum_{k=0}^{p} \frac{1}{(k+1)!} \, \mathrm{ad}_{\Omega_{(p+1)}}^k(\dot \Omega_{p+1}), \label{eq:omega_p+1_iter}
\end{align}
In the last sum,  the only term that can contribute at grade less than or equal to $p+1$ is the $k = 0$ term, that is
\begin{equation}
    \dot \Omega_{p+1},
\end{equation}
which has exactly grade $p+1$.

In the first two terms of \cref{eq:omega_p+1_iter}, terms where the nested commutators act only on $\Omega_{(p)}$ and $\dot \Omega_{(p)}$,together with $-A$, recover $\Theta_p$, while the remaining terms involve at least one occurrence of $\Omega_{p+1}$ and hence have grade at least $p+2$. More concretely, these additional terms take the form
\begin{equation}
    \ad_{Q_k}\cdots \ad_{Q_1}(\dot\Omega_{(p)}) 
\end{equation}
where at least one $Q_j$ is $\Omega_{p+1}$, ensuring the grade is at least $p+2$.

In summary, the only terms in $\Theta_{p+1}$ that can potentially contribute at grade $\leq p+1$ are:
\begin{equation} \label{eq:Theta_p_Theta_p+1}
    \Theta_p + \dot \Omega_{p+1}.
\end{equation}
By the inductive hypothesis, $\Theta_p$ contains no terms of grade $\leq p$.  
Since $\gd(\dot \Omega_{p+1}) = p+1$, it remains to verify that the grade $p+1$ terms in $\Theta_p + \dot \Omega_{p+1}$ cancel out.

The grade $p+1$ terms in $\Theta_p$ are
\begin{align}
  \ &   \sum_{k = 1}^p \frac{1}{(k+1)!} \sum_{j_0+\cdots +j_k = p+1 \atop j_0, \cdots, j_k \geq 1} 
    \ad_{\Omega_{j_k}}\cdots \ad_{\Omega_{j_1}} (\dot \Omega_{j_0})
\\
 = \ & \sum_{k = 1}^p \frac{1}{(k+1)!} \sum_{j_1+\cdots +j_k = p \atop j_1, \cdots, j_k \geq 1} 
    \ad_{\Omega_{j_k}}\cdots \ad_{\Omega_{j_1}} (A)
    + \sum_{k = 1}^p \frac{1}{(k+1)!} \sum_{j_0+\cdots +j_k = p+1 \atop j_1, \cdots, j_k \geq 1, j_0 \geq 2} 
    \ad_{\Omega_{j_k}}\cdots \ad_{\Omega_{j_1}} (\dot \Omega_{j_0})
    \\ 
     =:& I_1 + I_2,
\end{align}
where we split the terms according to $j_0 = 1$ and $j_0\geq 2$.
The grade $p+1$ terms in $\dot \Omega_{p+1}$ is itself, that is,
\begin{equation}\label{eq:dOmega_sum}
    \dot \Omega_{p+1} = \sum_{k=1}^{p} \frac{B_k}{k!}  S_{p+1}^{(k)}
    =  \sum_{k=1}^{p} \frac{B_k}{k!}
    \sum_{j_1+\cdots +j_k = p \atop j_1, \cdots, j_k \geq 1} 
    \ad_{\Omega_{j_k}}\cdots \ad_{\Omega_{j_1}} (A) = : I_3.
\end{equation}
By the definition of $S_{p+1}^{(k)}$, one can see that
\begin{equation}\label{eq:I2+I3}
   I_1 + I_3 = \sum_{k = 1}^p \left(\frac{1}{(k+1)!} + \frac{B_k}{k!} \right) S_{p+1}^{(k)}.
\end{equation}
We now focus on $I_2$. Further expanding $\dot \Omega_{j_0}$ using an equation analogous to \cref{eq:dOmega_sum} yields
\begin{align}\label{eq:I_2_sum}
    I_2 =& \sum_{k = 1}^p \frac{1}{(k+1)!} \sum_{j_0+\cdots +j_k = p+1 \atop j_1, \cdots, j_k \geq 1, j_0 \geq 2} 
    \ad_{\Omega_{j_k}}\cdots \ad_{\Omega_{j_1}} 
    \left( 
    \sum_{m=1}^{j_0-1} \frac{B_m}{m!}
    \sum_{\ell_1+\cdots +\ell_m = j_0-1 \atop \ell_1, \cdots, \ell_m \geq 1} 
    \ad_{\Omega_{\ell_m}}\cdots \ad_{\Omega_{\ell_1}} (A)
    \right)
    \\
    = & \sum_{k = 1}^p  \sum_{m=1, \cdots, p-1 \atop m+k \leq p} \frac{1}{(k+1)!}  \frac{B_m}{m!}
    \sum_{j_1 + \cdots j_k +  \ell_1+\cdots +\ell_m = p \atop 
    j_1, \cdots, j_k, \ell_1, \cdots, \ell_m \geq 1} 
     \ad_{\Omega_{j_k}}\cdots \ad_{\Omega_{j_1}} 
    \ad_{\Omega_{\ell_m}}\cdots \ad_{\Omega_{\ell_1}} (A)
    \\
    = & \sum_{k = 1}^p \sum_{m=1, \cdots, p-1 \atop m+k \leq p} \frac{1}{(k+1)!}  \frac{B_m}{m!} 
    S_{p+1}^{(m+k)}
    ,
\end{align}
where the second equality follows from reorganizing the summation indices in the first expression. More concretely, we take a closer look at the summation indices in the first line
\begin{equation}
    \sum_{k = 1}^p  
    \sum_{j_0+\cdots +j_k = p+1 \atop j_1, \cdots, j_k \geq 1, j_0 \geq 2}  \sum_{m=1}^{j_0-1}  
     \sum_{\ell_1+\cdots +\ell_m = j_0-1 \atop \ell_1, \cdots, \ell_m \geq 1}
     =    
     \sum_{k = 1}^p 
     \sum_{j_0 =2}^{p+1-k}
     \sum_{m=1}^{j_0-1}
    \sum_{j_1+\cdots +j_k = p+1-j_0 \atop j_1, \cdots, j_k \geq 1}    
     \sum_{\ell_1+\cdots +\ell_m = j_0-1 \atop \ell_1, \cdots, \ell_m \geq 1}.
\end{equation}
By reordering the summation over $j_0$ and $m$, we have
\begin{equation}   
     \sum_{k = 1}^p 
     \sum_{m=1}^{p-k}
     \sum_{j_0=m+1}^{p+1-k}
    \sum_{j_1+\cdots +j_k = p+1-j_0 \atop j_1, \cdots, j_k \geq 1}    
     \sum_{\ell_1+\cdots +\ell_m = j_0-1 \atop \ell_1, \cdots, \ell_m \geq 1}
     =      
     \sum_{k = 1}^p 
     \sum_{m=1}^{p-k} \sum_{j_1 + \cdots j_k +  \ell_1+\cdots +\ell_m = p \atop 
    j_1, \cdots, j_k, \ell_1, \cdots, \ell_m \geq 1},
\end{equation}
which recovers the summation indices in the second line of \cref{eq:I_2_sum}. We continue the investigation of $I_2$ by exchanging the summation order of $k$ and $m$ in the last line of \cref{eq:I_2_sum}.
\begin{align}
    I_2 = 
      \sum_{ m =1}^p \sum_{k =1}^{p-m} \frac{1}{(k+1)!}  \frac{B_m}{m!} 
    S_{p+1}^{(m+k)}
    = 
          \sum_{ m =1}^p \sum_{\tilde{k} =1+m}^{p} \frac{B_m}{m!(\tilde{k}-m+1)!}  
    S_{p+1}^{(\tilde{k})}.
\end{align}
Exchanging the summation order again yields
\begin{equation} \label{eq:I3}
    I_2 = 
       \sum_{\tilde{k} =2}^{p}
       \sum_{ m =1}^{\tilde{k}-1}
       \frac{B_m}{m!(\tilde{k}-m+1)!} 
    S_{p+1}^{(\tilde{k})}
    = -\sum_{\tilde{k} =2}^{p}
    \left(
    \frac{1}{\tilde{k} +1} +\frac{B_{\tilde{k}}}{\tilde{k}!}
     \right)    
    S_{p+1}^{(\tilde{k})},
\end{equation}
where we use the identity of Bernoulli numbers
\begin{equation}
    \sum_{m=0}^{\tilde{k}} \frac{(\tilde{k}+1)!}{m! (\tilde{k}+1-m)!} B_m
    = 
    \sum_{m=0}^{\tilde{k}} \binom{\tilde{k}+1}{m} B_m  = 0, \quad \tilde{k}\geq 1,
\end{equation}
so that
\begin{equation}
   \sum_{m=1}^{\tilde{k}-1} \frac{B_m}{m! (\tilde{k}+1-m)!}  =  
   - \frac{1}{(\tilde{k}+1)!}  
   - \frac{B_{\tilde{k}}}{\tilde{k}! },
\end{equation}
as $B_0 = 1$.
Combining \cref{eq:I2+I3} and \cref{eq:I3}, we arrive at
\begin{align}
   I_1 + I_2 + I_3 = & \sum_{k = 1}^p \left(\frac{1}{(k+1)!} + \frac{B_k}{k!} \right) S_{p+1}^{(k)} -\sum_{\tilde{k} =2}^{p}
    \left(
    \frac{1}{\tilde{k} +1} +\frac{B_{\tilde{k}}}{\tilde{k}!}
     \right)    
    S_{p+1}^{(\tilde{k})} 
    \\
    = & \left(\frac{1}{2} + B_1 \right)S_{p+1}^{(1)}  = 0,
\end{align} 
as $B_1 = -1/2$. This completes the proof.
\end{proof}
This establishes that $\Theta_p$ contains only terms of grade at least $p+1$. As a result, the leading error in $\tilde{A}_p - A$ is governed by nested commutators of grade $p+1$ or higher.
In this proof, we also obtain the following lemma as a byproduct (see the line after \cref{eq:Theta_p_Theta_p+1}).
\begin{lemma}\label{lem:gd_p+1_term_Theta_p_Omega_p+1}
    The grade $p+1$ terms in $\Theta_p$ matches those in $-\dot \Omega_{p+1}$.
\end{lemma}

\subsection{Error bounds with finite commutators}\label{sec:magnus_error_bounds}

Consider the coefficients
\begin{equation}\label{eq:def_c_pi_n}
c_{\pi, n} =  \frac{1}{n} (-1)^{d_b} \frac{d_a! \, d_b!}{n!},
\end{equation}
where $\pi \in S_n$ is a permutation, $d_a$ is the number of ascents in $\pi$, and $d_b$ is the number of descents. They satisfy $d_a + d_b = n-1$.
Each term in $\Omega_n$ is a grade $n$, width $n$ commutator associated with the coefficient $c_{\pi, n}$. To be more precise, for the permutation group $S_n$ with any positive interger $n$, the term $\Omega_n$ can be represented as
\begin{equation}\label{eqn:omega_k_expansion}
    \Omega_n(t) = \sum_{\pi\in S_n} c_{\pi,n} \int_{0}^{t} dt_1 \int_{0}^{t_1} dt_2 \cdots \int_{0}^{t_{n-1}} dt_{n} [A(t_{\pi(1)}),[A(t_{\pi(2)}),[\cdots, A(t_{\pi(n)})]],
\end{equation}
where $c_{\pi,n}$ is given by \cref{eq:def_c_pi_n}. See \cite[Section 2]{Arnal_2018} for a derivation of this fact. Such coefficients admit the following estimate.
\begin{lemma}\label{lem:bound_sum_c_pi_n}
    \begin{equation}
        \sum_{\pi \in S_n} |c_{\pi, n}|  \leq \frac{(n-1)!}{n}.
    \end{equation}
\end{lemma}
\begin{proof}
    The maximum of $d_a! d_b!$ under the condition $d_a + d_b = n-1$ is bounded by $(n-1)!$. We therefore have
    \begin{equation}
        \sum_{\pi \in S_n} |c_{\pi, n}| \leq n! \cdot \frac{1}{n^2} = \frac{(n-1)!}{n}.
    \end{equation}
\end{proof}

Thus, each $\Omega_n$ is a grade-$n$, width-$n$ term with the total coefficient sum bounded by $\frac{(n-1)!}{n}$.
Using this bound on the sum of coefficients, we have for each 
$\Omega_n$:
\begin{equation}\label{eq:norm_Omega_n}
\norm{\Omega_n} \leq \frac{(n-1)!}{n}  \beta_{\mathrm{comm},n} \leq \frac{1}{n^2} \alpha_{\mathrm{comm},n} t^n,
\end{equation}
where $\beta_{\mathrm{comm},n}$ denotes the maximal norm among all grade-$n$, width-$n$ commutator terms. 
For simplicity, we omit the explicit $t$-dependence in $\beta_{\mathrm{comm},n}$. 
However, because of the time-ordered integrals, $\beta_{\mathrm{comm},n}$ can be bounded above by $\alpha_{\mathrm{comm},n} t^n / n!$, where $\alpha_{\mathrm{comm},n}$ is the largest norm of any grade-$n$ nested commutator of $A$ with no time integration. 
We also have that $\dot \Omega_n$ is a grade $n$ and width $n-1$ term with the sum of its coefficients upper bounded by $\frac{(n-1)!}{n}$. In particular,
\begin{equation}\label{eq:norm_dOmega_n}
    \norm{\dot \Omega_n} \leq \frac{(n-1)!}{n} \gamma_{\mathrm{comm},n} \leq \frac{1}{n}\alpha_{\mathrm{comm},n} t^{n-1},
\end{equation}
where $\gamma_{\mathrm{comm},n}$ denotes the maximal norm among all grade-$n$, width-$n-1$ commutator terms, which is upper bounded by  $\alpha_{\mathrm{comm},n} t^{n-1} / (n-1)!$.

By \cref{lem:gd_p+1_term_Theta_p_Omega_p+1}, we have the upper bound of the grade $p+1$ terms in $\Theta_p$, which are thus upper bounded by 
\begin{equation}
    \frac{1}{p+1}\alpha_{\mathrm{comm},p+1} t^{p}.
\end{equation}
We now turn to the $p+2$ grade terms in $\Theta_p$. We start by counting the coefficients of the grade $p+2$ terms in $\Theta_p$. Recall that 
\begin{equation}\label{eq:recall_theta_p}
  \Theta_p =    \sum_{k=0}^{p} \frac{1}{(k+1)!} \,
   \mathrm{ad}_{\Omega_{(p)}}^k(\dot \Omega_{(p)}) - A,
\end{equation}
and we can expand $\Omega_{(p)}$ and $\dot \Omega_{(p)}$ with each of their component upper bounded by \cref{eq:norm_Omega_n} and \cref{eq:norm_dOmega_n}, and collect all grade $p+2$ terms -- they all have width $p+1$. As $k = 0$ term in $\Theta_p$ all have grade smaller than $p$, we can consider the index in the sum from $k = 1$. Note that the expression of $\Theta_p$ as given in~\cref{eq:recall_theta_p} includes $\operatorname{ad}^k$ (commuting $k$ times). To collect all contributions of overall grade $p+2$, we consider components $n_1,\dots,n_k$ satisfying $n_1 + \cdots + n_k = p+2$. Each grade $n_j$ term is associated with a coefficient upper bounded by $(n_j-1)!/n_j = n_j!/n_j^2$ (as shown in~\cref{lem:bound_sum_c_pi_n}). Hence, the total contribution from all grade $(p+2)$ terms can be estimated by
\begin{align}\label{eq:gd_p+1_counting_estimate_in_Theta_p}
    & \sum_{k=1}^p \frac{1}{(k+1)!} \sum_{\substack{n_0 + n_1 + \cdots + n_k = p+2 \\ n_0, \cdots, n_k \geq 1}} \frac{n_0! \, n_1! \cdots n_k!}{n_0^2 n_1^2 \cdots n_k^2} \gamma_{\mathrm{comm},p+2} 
    \\
    \leq & 
\sum_{k=1}^p \frac{1}{(k+1)!} \sum_{\substack{n_0 + n_1 + \cdots + n_k = p+2 \\ n_0, \cdots, n_k \geq 1}} \frac{(n_0-1)! \, (n_1-1)! \cdots (n_k-1)!}{n_0 n_1 \cdots n_k} \gamma_{\mathrm{comm},p+2} 
      \\
    \leq & 
    \sum_{k=1}^p \frac{1}{(k+1)!} \frac{(p+1-k)!}{2}  \binom{p+1}{k} \gamma_{\mathrm{comm},p+2}, 
\end{align}
where we used the fact that 
\begin{equation}
    (n_0-1)! \, (n_1-1)! \cdots (n_k-1)! \leq (p+1-k)! , \quad n_0 n_1 \cdots n_k \geq 2, 
\end{equation}
for $n_0+\cdots +n_k = p+2$ with $k \leq p$ and $n_0,\cdots, n_k \geq 1$, and there are $ \binom{p+1}{k} $ terms in the sum of $n_0, \cdots, n_k$. 
So the norm of the grade $p+2$ terms is upper bounded by 
\begin{align}
     & \sum_{k=1}^p \frac{1}{(k+1)!} \frac{(p+1-k)!}{2}  \binom{p+1}{k} \gamma_{\mathrm{comm},p+2}
     \\
     = & 
     \sum_{k=1}^p \frac{1}{(k+1)!} \frac{(p+1-k)!}{2}  \frac{(p+1)!}{k!(p+1-k)!} \gamma_{\mathrm{comm},p+2}
     \\
     \leq &  \frac{1}{2}(p+1)!\sum_{k=1}^p\frac{1}{(k+1)! k!}  \gamma_{\mathrm{comm},p+2}
     \leq (p+1)!  \gamma_{\mathrm{comm},p+2},
\end{align}
where we used the inequality that
\begin{equation}
    \sum_{k=1}^p\frac{1}{(k+1)! k!}  \leq \sum_{k=1}^\infty\frac{1}{k^2}  = \frac{\pi^2}{6} \leq 2.
\end{equation}
Similarly, for the grade $p+3, p+4, \cdots$. We have for any $q\geq p+2$, the grade $q$ terms in $\Theta_p$ is upper bounded by
\begin{equation}
 (q-1)!  \gamma_{\mathrm{comm},q},
\end{equation}
which can be further bounded by 
\begin{equation}
 (q-1)!  \frac{t^{q-1}}{(q-1)!}\alpha_{\mathrm{comm},q} = t^{q-1}\alpha_{\mathrm{comm},q},
\end{equation}
as the grade $q$ term has width $q-1$ leading to a factor of $t^{q-1}/((q-1)!)$ from the time integral.

Therefore, we have
\begin{equation}\label{eq:theta_p_norm_in_comm}
    \norm{\Theta_p} \leq  \frac{1}{p+1}\alpha_{\mathrm{comm},p+1} t^{p} +  \sum_{q=p+2}^{p^2 + p}t^{q-1}\alpha_{\mathrm{comm},q}.
\end{equation}

For the remainder term associated with $g_p(\cdot)$ in \cref{eq:tildeA_p-A_def}, it can be controlled via terms involving the time-integral of the nested commutator of $A$ with grade $\geq p+2$ and width $\geq p+1$. 
\begin{lemma} \label{lem:g_remainder}
For $\Omega_{(p)}$ and $g_p$ defined in \cref{eq:mag_p_unitary_t_only} and \cref{eq:def_g_p}, we have 
\begin{equation}
    \norm{g_p(\ad_{\Omega_{(p)}})(\ad_{\Omega_{(p)}}^{p+1}(\dot \Omega_{(p)}) )}\leq C \norm{\ad_{\Omega_{(p)}}^{p+1}(\dot \Omega_{(p)})},
\end{equation}
where $C$ is some absolute constant.
\end{lemma}
The proof is essentially the same as \cite[Lemma 7]{FangLiuSarkar2025} and \cite[Lemma 5.1]{HochbruckLubich2003}, which we include in~\cref{sec:appendix_pf_g_remainder} for completeness. 

We now consider the upper bound of 
\begin{equation}\label{eq:remainder_wo_g}
   \mathrm{ad}_{\Omega_{(p)}}^{p+1} (\dot \Omega_{(p)} ),
\end{equation}
which contains terms with their grade at least $p+2$. Following a similar counting estimate as in \cref{eq:gd_p+1_counting_estimate_in_Theta_p}, the grade $q$ ($q\geq p+2$) term in \cref{eq:remainder_wo_g} can be upper bounded by
\begin{align}
   & \,  
   \sum_{\substack{n_0 + n_1 + \cdots + n_{p+1} = q \\ n_0, \cdots, n_{p+1} \geq 1}} \frac{n_0! \, n_1! \cdots n_{p+1}!}{n_0^2 n_1^2 \cdots n_{p+1}^2} \gamma_{\mathrm{comm},q} 
    \\
    \leq & \,
    (q-p-2)! \binom{q-1}{p+1} \gamma_{\mathrm{comm},q} 
     \\
    \leq & \,
    (q-p-2)! \binom{q-1}{p+1} \frac{t^{q-1}}{(q-1)!}\alpha_{\mathrm{comm},q} 
   =
   \frac{t^{q-1}}{ (p+1)!} \alpha_{\mathrm{comm},q}. 
\end{align}
Combining terms in all grade $q \geq p+2$, we have the upper bound of \cref{eq:remainder_wo_g} as
\begin{equation}\label{eq:g_part_norm_in_comm}
 \sum_{q=p+2}^{p^2+2p} \frac{t^{q-1}}{ (p+1)!} \alpha_{\mathrm{comm},q}.
\end{equation}
Combining \cref{eq:g_part_norm_in_comm} with \cref{eq:theta_p_norm_in_comm}, we obtain an upper bound of $\tilde{A}-A$. Multiplying this bound by $t$ yields the local truncation error due to \cref{eq:U-Up_integral_in_tildeA-A}:
\begin{equation}
    \frac{1}{p+1}\alpha_{\mathrm{comm},p+1} t^{p+1} +  \sum_{q=p+2}^{p^2+p}t^{q}\alpha_{\mathrm{comm},q} +  C\sum_{q=p+2}^{p^2+2p} \frac{t^{q}}{ (p+1)!} \alpha_{\mathrm{comm},q},
\end{equation}
where $C$ is an absolute constant.
We can combine the second and third terms for a more streamlined expression. This leads to the following local truncation error estimate for a short time $t$.
\begin{thm}[Short-time error of $p$-th order Magnus expansion]\label{thm: local_magnus_error}
Let $t > 0$, and let $U_p(t)$ denote the $p$-th order Magnus approximation as defined in \cref{eq:mag_p_unitary_t_only}, while $U(t)$ denotes the exact evolution operator as given in \cref{eq:def_U(t)} over the interval $[0,t]$. Then the error of the $p$-th order Magnus expansion depends only on finitely many nested commutators of grade at least $p+1$, and satisfies
\begin{equation}
\norm{U_p(t) - U(t)} \leq \frac{1}{p+1}\alpha_{\mathrm{comm},p+1} t^{p+1}
+ C \sum_{q=p+2}^{p^2+ 2p} \alpha_{\mathrm{comm},q} t^{q},
\end{equation}  
where $C$ is an absolute constant.
\end{thm}
 
We remark that the estimate in the second term, which involves orders of $t$ higher than $p+2$, may not be tight. Nonetheless, it already provides the correct asymptotic scaling of the quantum algorithm as desired.

Although \cref{thm: local_magnus_error} holds for any $t > 0$, we refer to it as the short-time error since, in our algorithm, the $p$-th order Magnus expansion is applied only over short intervals $[t_j, t_{j+1}]$ of length $h = t_{j+1} - t_j$. For the local truncation error on such an interval, the argument proceeds exactly as before, except that the time integrals are taken over $[t_j, t_{j+1}] = [t_j, t_j + h]$ rather than $[0,t]$.

\begin{thm}[Global error of $p$-th order Magnus expansion] \label{thm:magnus_trunction_finite_comm}
 Let $U(T,0) = \mathcal{T}e^{-i\int_0^T H(s) \, ds}$ be the exact long-time propagator over the time interval $[0, T]$, and let $U_p(T,0)$ denote the approximation obtained by applying the $p$-th order Magnus expansion with time step size $h = T/L$, i.e.,
\begin{equation}
    U_p(t_L,t_{L-1})\dots U_p(t_2,t_1)U_p(t_1,t_0).
\end{equation}
where $t_j = j h$ and each $U_p(t_{j+1}, t_j)$ is the $p$-th order Magnus approximation over the subinterval.
We have the following results regarding the long-time error between them:
\begin{enumerate}
\item Let $\alpha_{\mathrm{comm},q}$ denote the maximum norm (taken over time and over all nested commutators of grade $q$) of the operator $H(t)$. That is,
 \begin{equation}\label{eq:def_alpha_comm_q}
\alpha_{\mathrm{comm},q} := \sup_{\tau_1, \ldots, \tau_q \in [0,T]} \max_{C \in \mathcal{C}_q} \big\| C\big(H(\tau_1), \ldots, H(\tau_q)\big) \big\|,
 \end{equation}
where $\mathcal{C}_q$ denotes the set of all nested commutators of grade $q$. 
Then the global error is bounded by
\begin{equation}
  \norm{U(T,0) - U_p(T,0)} \leq   
  \frac{1}{p+1}\alpha_{\mathrm{comm},p+1}  h^{p} T + C \sum_{q = p+2}^{p^2+2p} \alpha_{\mathrm{comm},q}  h^{q-1} T,
\end{equation}
for some absolute constant $C > 0$.

\item Define
\begin{equation}\label{eq:def_bar_alpha_comm}
\bar \alpha_\mathrm{comm} := \max_{p+1 \leq q \leq p^2+2p} \alpha_{\mathrm{comm}, q}^{1/q}.
\end{equation}
To ensure that the global error is bounded by a target precision $\epsilon$,
it suffices to choose parameters such that
\begin{equation}\label{eqn: parameters_for_optimal_HAMT}
\bar \alpha_\mathrm{comm}  h = \Theta(1), \quad L = \Or(\bar \alpha_\mathrm{comm} T),
\quad
p = \mathcal{O}\left( \log\left( \frac{\bar \alpha_\mathrm{comm} T}{\epsilon} \right) \right),
\end{equation}
or 
for a fixed integer $p$, we can choose
\begin{equation}
      L = \mathcal{O}{ \left( \frac{\bar \alpha_{\mathrm{comm}}^{1+1/p} T^{1+1/p}}{ \epsilon^{1/p}} \right)}.
\end{equation}
\end{enumerate}

\end{thm}
\begin{rem}
We revealed that the error bound for $p$-th order Magnus expansion depends only on the nested commutator of $A(\cdot)$ with grade $\geq p+1$. It is also possible to make the error in terms of $L^1$ norm scaling in terms of $\int_0^t \norm{A(s)}\, ds$. This is because in the error representation, each grade $q+1$ term has width $q$, i.e. $q$ layers of integrals, and together with
\begin{equation}
 \bar \alpha_\mathrm{comm} \leq 2 \sup_{s \in [0,t]}\norm{A(s)} \leq 2\alpha,
\end{equation}
this yields the desired $L^1$ scaling.
\end{rem}

\begin{rem}\label{rem:infinite_sum}
We also remark that our approach reveals that the truncation error depends only on a finite number of nested commutators. Alternatively, one can demonstrate the commutator scaling -- when focusing solely on nested commutators of grade larger than $p+1$ -- by directly using \cref{eq:norm_Omega_n}, which can easily lead to the local and global errors which we include as \cref{thm:magnus_trunction_infty_comm} in \cref{sec:appendix_pf_g_remainder} as it is not used in our work.
The key difference between \cref{thm:magnus_trunction_finite_comm} and \cref{thm:magnus_trunction_infty_comm} lies in the number of nested commutators involved: \cref{thm:magnus_trunction_finite_comm} depends only on a finite number, whereas \cref{thm:magnus_trunction_infty_comm} involves infinitely many. The latter kind of commutator form is typically used to achieve logarithmic scaling in precision, such as in the multiproduct formula~\cite{LowKliuchnikovWiebe2019,AftabAnTrivisa2024,MizutaIkedaFujii2024}. In our setting, we prove that the error depends on only finitely many commutators.
\end{rem}

\begin{rem}
We remark that there are two main differences compared to the commutator scaling in the generalized Trotterization~\cite[Corollary 11]{MizutaIkedaFujii2024}. First, Trotter-Suzuki involves a constant $C = V^p$, where $V$ denotes the number of layers included in the time-dependent product formula which is given by $2 \cdot 5^{p/2-1}$, while ours does not have such $p$-dependent factors. Second, the Magnus commutators are expressed in terms of the Hamiltonian itself and do not rely on its derivatives.
\end{rem}

\section{Quadrature error of the $p$-th order Magnus expansion}\label{sec: dis_error}
In this section, we analyze the error introduced by discretizing the time-ordered integrals in the Magnus expansion using numerical quadrature. 
We first present the explicit form of the discretized Magnus terms and then derive bounds for the local quadrature error associated with each term. 
We then establish a general error bound for the entire $p$-th order Magnus expansion when the integrals are approximated by quadrature rules.

Recall that the $p$-th order Magnus expansion can be denoted as
\begin{equation}
    \Omega_{(p)}(t,0) = \sum_{k=1}^{p} \Omega_k(t,0),
\end{equation}
where one can expand the commutators in each $\Omega_k$ and rewrite each term as
\begin{equation}
    \Omega_k(t,0) = \sum_{\pi\in S_k} C_{\pi,k} \int_{0}^{t} dt_1 \int_{0}^{t_1} dt_2 \cdots \int_{0}^{t_{k-1}} dt_{k} A(t_{\pi(1)})A(t_{\pi(2)})\cdots A(t_{\pi(k)}).
\end{equation}
Here the coefficients $C_{\pi,k}$ admit an explicit expression as follows.
\begin{lemma}
\label{lem:gold}
The coefficient $C_{\pi,k}$ equals
\begin{equation}
    C _{\pi,k}= \frac{(-1)^{d_a(\pi)}}{k} \times \frac{1}{ {k-1 \choose d_a(\pi)}}
\end{equation}
where $d_a(\pi)=\left|\{\,i\in\{1,\dots,k-1\}|\pi(i)>\pi(i+1)\}\right|$ denotes the number of descents of permutation $\pi$ and $k$ denotes the length of permutation.
\end{lemma}
\begin{proof}
See proof at~\cite[Sec.~$2$]{Arnal_2018}.
\end{proof}

The discretized Magnus expansion $\tilde{\Omega}_{(p)}(t,0)$ can be defined as
\begin{equation}
\label{equ:kquadrature}
    \tilde{\Omega}_{(p)}(t,0) = \sum_{k=1}^{p} \tilde{\Omega}_k(t,0),
\end{equation}
where 
\begin{equation}
    \tilde{\Omega}_k(t,0) = \sum_{\pi \in S_k} C_{\pi,k} \sum_{i_1=0}^{M-1} \sum_{i_2=0}^{i_1-1}\dots \sum_{i_k=0}^{i_{k-1}-1}  \ \prod_{l=1}^{k} A\left(\frac{i_{\pi(l)}t}{M}\right) \ \frac{t^k}{M^k}. 
\end{equation}

\begin{lemma}[Local quadrature error of multi-layer time-ordered integrals]
\label{lem:quadrature}

Let $t > 0$ and let $M$ be the number of time steps used to discretize each time integral in the $k$-th order Magnus term. Then the quadrature error for approximating $\Omega_k(t,0)$ by its discretized version $\tilde{\Omega}_k(t,0)$ satisfies

\begin{equation}
    \left\| \tilde{\Omega}_k(t,0)-\Omega_{k}(t,0)\right\| \leq \frac{1}{M}\left( k t^{k+1}  \|A'\| \|A\|^{k-1} + (k-1) t^k \|A\|^k\right).
\end{equation}

\end{lemma}
\begin{proof}
To simplify notation, we define
    \begin{equation}
        \Delta_{k,\sigma} := \{ (t_1,t_2,\dots,t_{k}) \in \mathbb{R}^{k} | \  0\leq t_{\sigma(1)}\leq t_{\sigma(2)}\leq \cdots \leq t_{\sigma(k)} \leq t \ \},
    \end{equation}
    where $\sigma$ is an element in the $k$-th order permutation group $S_k$ and then $\Omega_k(t,0)$ can be represented as
    \begin{equation}
        \sum_{\pi\in S_k} C_{\pi,k} \int_{\Delta_{k, \sigma_0}} A(t_{\pi(1)})\cdots A(t_{\pi(k)}) \ d t_1\dots d t_k
    \end{equation}
    where $\sigma_0(i)=i$ for any $i\in \{1,2,3,\dots, k\}$.
    To analyze the local quadrature error of the $k$-th order Magnus expansion term, we recall that the discretized $\tilde{\Omega}_k(t,0)$ defined as
   \begin{equation}
     \sum_{\pi \in S_k} C_{\pi,k} \sum_{i_1=0}^{M-1} \sum_{i_2=0}^{i_1-1}\dots \sum_{i_k=0}^{i_{k-1}-1}  \ \prod_{l=1}^{k} A\left(\frac{i_{\pi(l)}t}{M}\right) \ \frac{t^k}{M^k},
\end{equation}
    can be viewed as the integral of a piecewise constant function over a convex domain within $\mathbb{R}^k$, denoted by $\tilde{\Delta}_{k}$. It consists of all the $k$-dimensional cubes contained in $\Delta_{k,\sigma_0}$ each of side length $\frac{t}{M}$. Since there are some cubes only partially contained within $\Delta_{k,\sigma_0}$, it can be seen that
    \begin{equation}
        \tilde{\Delta}_k \subset \Delta_{k}.
    \end{equation}

    To rigorously estimate the $\|\tilde{\Omega}_k-\Omega_k\|$, we need to estimate two sources leading to the error separately. Note that
    \begin{equation}
    \label{equ:omega2}
        \begin{split}
            \Omega_k-\tilde{\Omega}_k &= \sum_{\pi \in S_k} C_{\pi,k} \iiint \dots \int_{\tilde{\Delta}_{k,\sigma_0}}  dt_1\dots dt_k \left (  A(t_{\pi(1)})\cdots A(t_{\pi(k)}) - \tilde{A}(t_{\pi_{(1)}})\dots \tilde{A}(t_{\pi(k)}) \right)\\
            +& \sum_{\pi \in S_k} C_{\pi,k} \iiint \dots \int_{\partial \tilde{\Delta}_{k,\sigma_0}} dt_1\dots dt_k A(t_{\pi(1)})\cdots A(t_{\pi(k)})
        \end{split}
    \end{equation}
    where $\tilde{A}$ is a piecewise constant function leading to the numerical quadrature~\cref{equ:kquadrature} and the boundary $\partial \tilde{\Delta}_{k,\sigma} := \Delta_{k,\sigma} \setminus \tilde{\Delta}_{k,\sigma}$ for any $\sigma \in S_{k}$. 
    
    Since $A$ is differentiable, by the midpoint rule, we have that
    \begin{equation}
        \left \|   A(t_{\pi(1)})\cdots A(t_{\pi(k)}) - \tilde{A}(t_{\pi_{(1)}})\dots \tilde{A}(t_{\pi(k)}) \right\| \leq \frac{kt}{M} \|A' \|\| A\|^{k-1}.
    \end{equation}
    Therefore, the first term in \cref{equ:omega2} can be bounded by
    \begin{equation}
    \begin{split}
         &\left \| \sum_{\pi \in S_k} C_{\pi,k} \iiint \dots \int_{\tilde{\Delta}_k}  dt_1\dots dt_k k \|A'\| \|A\|^{k-1} \frac{t}{M}\right \|\\
         =& \left \| \sum_{\pi \in S_k} C_{\pi,k} \frac{kt^k}{k!} \|A'\| \|A\|^{k-1} \frac{t}{M}\right \| \leq \frac{1}{M}k t^{k+1}  \|A'\| \|A\|^{k-1}.
    \end{split}
    \end{equation}
The last inequality comes from the fact that $|C_{\pi,k}|\leq 1$.

For the second term in \cref{equ:omega2} involving the boundary $\partial \tilde{\Delta}_k$, we aim to show that 

\begin{equation}
    \sum_{\pi \in S_k} C_{\pi,k} \iiint \dots \int_{\partial \tilde{\Delta}_k} \mathrm{d}t_1\dots \mathrm{d}t_k A(t_{\pi(1)})A(t_{\pi(2)})\cdots A(t_{\pi(k)}) \leq\sum_{\pi \in S_k}  \frac{C_{\pi,k}}{k!}  \frac{ \|  A\|^k t^k k}{2M},
\end{equation}
by estimating the volume of $\tilde{\Delta}_{k,\sigma_0}$.
Furthermore, since $|C_{\pi,k}|\leq 1$, the integral is bounded by $\frac{ \|  A\|^k t^k k}{2M}$. In order to establish the estimate, we claim that

\begin{equation}
    \cup_{\sigma\in S_k} \partial_{\tilde{\Delta}_{k,\sigma}} = [0,t]^{k}-\cup_{\sigma\in S_k} \tilde{\Delta}_{k,\sigma}.
\end{equation}
This follows from the definition of $\partial_{\tilde{\Delta}_{k,\sigma}}$ and $\tilde{\Delta}_{k,\sigma}$. 
In addition, we note that for any $\sigma \in S_k$, the permutation $\sigma$ acts as an isomorphism such that
\begin{equation}
    \sigma(  \Delta_{k,\sigma}) = \Delta_{k,\sigma_0},
\end{equation}
and
\begin{equation}
    \sigma([0,t]^k)=[0,t]^k.
\end{equation}
Furthermore, these properties generalize to $\tilde{\Delta}_{k,\sigma}$ and $\partial\tilde{\Delta}_{k,\sigma}$. To be specific,

\begin{equation}
    \sigma(\tilde{\Delta}_{k,\sigma}) = \tilde{\Delta}_{k,\sigma_0}
\end{equation}
and 

\begin{equation}
    \sigma(\partial\tilde{\Delta}_{k,\sigma}) = \partial\tilde{\Delta}_{k,\sigma_0}.
\end{equation}
Moreover, for any $\sigma\neq\sigma'$, we have that
\begin{equation}
    \label{equ:meausre}\mu(\partial\tilde{\Delta}_{k,\sigma} \cap \partial\tilde{\Delta}_{k,\sigma'})= 0 \quad  \text{and} \quad \mu(\tilde{\Delta}_{k,\sigma} \cap \tilde{\Delta}_{k,\sigma'})= 0,
\end{equation}
where $\mu$ is used to denote Lebesgue measure. Indeed, use $\text{int}(S)$ to denote the interior of the set S and then we have

\begin{equation}
    \text{int} (\ \partial\tilde{\Delta}_{k,\sigma} \cap \text{int} \ \partial\tilde{\Delta}_{k,\sigma'}) = \varnothing.
\end{equation}
Therefore, we obtain
\begin{equation}
\label{equ:euqal}    \mu(\partial\tilde{\Delta}_{k,\sigma_0})=\mu\left(\frac{1}{k!}\cup_{\sigma}\partial\tilde{\Delta}_{k,\sigma}\right),
\end{equation}
and $\partial\tilde{\Delta}_{k,\sigma_0}$ almost equal to $\frac{1}{k!}\cup_{\sigma}\partial\tilde{\Delta}_{k,\sigma}$ with difference of measure zero.

Next, we estimate the volume of the boundary $\frac{1}{k!}\cup_{\sigma}\partial\tilde{\Delta}_{k,\sigma}$. 
As a direct consequence of \cref{equ:meausre} , the union of all such boundaries corresponds exactly to the $k$-dimensional cubes, which intersect with hyperplanes imposing the ordering conditions 
\begin{equation}
    t_1=t_2, \quad t_2=t_3,  \quad \dots, \quad t_{k-1}=t_{k},
\end{equation}
together with the hyperplanes defining the range of points, namely $t_i=0$ and $t_i=t$ for any $i\in \{1,2,3\dots,k \}$. 
If a $k$-dimensional cube does not intersect any boundary non-trivially (the Lebesgue measure of the intersection is nonzero), then it must lie in a interior region $\text{int} (\partial \tilde{\Delta}_{k,\sigma^{*}})$ for a $\sigma^{*}\in S_{k}$. 
Therefore, any cube $\square_k$ that intersects at least one hyperplane $t_i=t_{i+1}$ belongs to the union of all boundaries $\cup_{\sigma}\partial\tilde{\Delta}_{k,\sigma}$.

To derive an upper bound for the volume of $\cup_{\sigma}\partial\tilde{\Delta}_{k,\sigma}$, we first note that projecting $\square_k$ onto the $(t_i,t_{i+1})$-plane (with all other coordinates set to zero) yields a two-dimensional cube that intersects the hyperplane $t_i = t_{i+1}$. 
Therefore, we can show that the projection of any cube $\square_k$ intersecting $t_i=t_{i+1}$ has area
\begin{equation}
    \frac{t^2}{M^2}. 
\end{equation}
As a result, the volume of all boundary cubes that intersect with the hyperplane $t_i=t_{i+1}$ is at most
\begin{equation}
    M\cdot\frac{t^2}{M^2} \cdot t^{k-2}.
\end{equation}
Since there are $k-1$ different hyperplanes, the total volume of boundary terms can be bounded by
\begin{equation}
    \left| \cup_{\sigma}\partial\tilde{\Delta}_{k,\sigma} \right|\leq\frac{(k-1)t^2}{M} \cdot t^{k-2}.
\end{equation}
By \cref{equ:euqal}, it follows that
\begin{equation}
    \left | \partial\tilde{\Delta}_{k,\sigma_0} \right |\leq \frac{(k-1)t^2}{k!M} \cdot t^{k-2}.
\end{equation}
\end{proof}

The following theorem summarizes the local quadrature error for the $p$-th order Magnus series.
\begin{thm}[Local quadrature error of $p$-th order Magnus expansion] \label{thm: local_quadrature_error}
Let $t>0$, and discretize each time integral in $\tilde{\Omega}_{(p)}$ using $M$ time steps. Then  

\begin{equation}
    \norm{ \tilde{\Omega}_{(p)}(t,0)-\Omega_{(p)}(t,0)} \leq \sum_{k=1}^{p}\frac{1}{M}\left( k t^{k+1}  \|A'\| \|A\|^{k-1} + (k-1) t^k \|A\|^k\right). 
\end{equation}
\end{thm}
\begin{proof}
The estimate follows directly from \cref{lem:quadrature}.
\end{proof}

\section{Circuit construction} \label{sec: circuit}

In this section, we present the construction of quantum circuits for simulating quantum dynamics using the truncated Magnus expansion. 
We first describe the quantum input model for time-dependent Hamiltonians in \cref{sec:input}.
We then implement the Magnus expansion algorithm on quantum circuits in \cref{sec: quantum_circuit_magnus}, where we construct block encodings of the discretized Magnus terms $\tilde\Omega_k$, and discuss how these are combined into $\tilde \Omega_{(p)}$ using linear combinations of unitaries techniques.

\subsection{Input model}\label{sec:input}

The input model (HAM-T oracle) for a time-dependent Hamiltonian $H(t)$ with $\|H(t)\|\leq \alpha$ is defined as follows.

\begin{defn}[HAM-T Oracle]
Let $M$ be the number of quadrature points used in the Magnus expansion at each local time step, let $j$ index the local time step, and let $h$ be the time-discretization step size. 
We define $\text{HAM-T}_j$ as an $(n_s+n_a+n_m)$-qubit unitary, where $n_m = \log M$, such that
\begin{equation}
    \bra{0^{n_a}} \text{HAM-T}_j \ket{0^{n_a}}
    = \sum_{k=0}^{M-1} \ket{k}\bra{k} \otimes \frac{H(jh + kh/M)}{\alpha}.
\end{equation}
Here $n_s, n_a$ denote the numbers of state-space and ancilla qubits, respectively.
\end{defn}

We note that in the following we use an HAM-T oracle to encode $A=-iH$ for notational simplicity, which is essentially equivalent to the definition.
This is a commonly used time-dependent matrix encoding for time-dependent Hamiltonian simulation as proposed in~\cite{LowWiebe2019}. The unitary oracle HAM-T encodes the time-dependent Hamiltonian at discrete time steps. The construction of the HAM-T oracle can be reduced to block encoding of Hamiltonian at each time step~\cite{GilyenSuLowEtAl2019,sunderhauf2024block}. 
This encoding for a number of Hamiltonians, including sparse matrices~\cite{camps2024explicit}, pseudo-differential operators~\cite{li2023efficient}, and quantum many-body Hamiltonians~\cite{babbush2018encoding,Wan2021exponentially,liu2024efficient,du2024hamiltonian,liu2025block} have been constructed explicitly.

\subsection{Quantum circuit for Magnus expansion}\label{sec: quantum_circuit_magnus}
We recall the $p$-th order Magnus expansion (\cref{eq:mag_p_unitary_tj_tj+1}) given as
\begin{equation}
    \Omega_{(p)}(t_j+h,t_j) = \sum_{k=1}^{p} \Omega_k(t_j+h,t_j),
\end{equation}
where $t_j=jh$ and $h=\frac{T}{L}$ denotes the time step. The corresponding evolution operator is defined as 
\begin{equation}
    U_p(t_j+h, t_j) = e^{\Omega_{(p)}(t_j+h,t_j)}.
\end{equation}
Here, we note that $\Omega_k$ can be represented as a weighted sum of $k$-layer time integrals of $k$-layer, right-normed nested commutators. 
Recall that an alternative representation of the Magnus expansion can be derived from expanding the right-normed nested commutators:
\begin{equation}
\label{equ:high_order_term}
    \Omega_k(t,0) = \sum_{\pi\in S_k} C_{\pi,k} \int_{0}^{t} dt_1 \int_{0}^{t_1} dt_2 \cdots \int_{0}^{t_{k-1}} dt_{k} A(t_{\pi(1)})A(t_{\pi(2)})\cdots A(t_{\pi(k)}),
\end{equation}
and $C_{\pi,k}$ has an explicit representation as given in \cref{lem:gold}.

\subsubsection{State preparation for time-ordered multi-layer integral}

In this subsection, we discuss state preparation for discretized time-ordered integrals, a key subroutine in constructing the block-encoding circuit for higher-order Magnus expansion. As illustrated in the high-order Magnus expansion~\cref{equ:high_order_term}, each multi-layer integral within the sum requires a decreasing ordering of the times $t_1,t_2,\dots, t_k$. Therefore, the discretized time-ordered integral must satisfy a similar condition.
For a given multi-layer integral term in $\Omega_{k}(t_j+h,t_j)$, its discretization can be written as
\begin{equation}\label{eqn: tilde_omega_k}
\begin{split}
      \sum_{i_1=0}^{M-1} \sum_{i_2=0}^{i_1-1} \sum_{i_3=0}^{i_2-1} \dots \sum_{i_k=0}^{i_{k-1}-1} C_{\pi,k}\, A\left(t_j+\frac{ i_{\pi(1)}h}{M} \right)A\left( t_j+\frac{i_{\pi(2)}h}{M} \right) 
     A\left( t_j + \frac{i_{\pi(3)}h}{M}\right) \dots A\left( t_j + \frac{i_{\pi(k)}h}{M}\right) \frac{h^k}{M^k}.
\end{split}
\end{equation}
It can be seen that the summation requires $M-1\geq i_1>i_2>\dots >i_k \geq 0$. 
To simplify notation, we introduce an alternative expression for the sum,
\begin{equation}
\sum_{i_1>i_2>\dots>i_k}^{}:=\sum_{i_1=0}^{M-1} \sum_{i_2=0}^{i_1-1} \sum_{i_3=0}^{i_2-1} \dots \sum_{i_k=0}^{i_{k-1}-1}.
\end{equation}

To construct the block encoding of the discretized integral with $\mathcal{O}(k)$ query complexity in HAM-T, we aim to construct a prepare oracle that outputs
\begin{equation}
   \sqrt{\frac{k!}{M^k}} \sum_{i_1>i_2>\dots>i_k}\ket{i_1}\ket{i_2} \dots \ket{i_k}. 
\end{equation}

However, since the state is not normalized, we cannot prepare it exactly. 
Instead, we prepare the state using post-selection with the methods developed in~\cite{berry2018improved}. 
We refer the reader to~\cite{berry2018improved} for the detailed construction, and state here only the version needed for this paper, together with a sketch of the proof. 
Note that in~\cite{berry2018improved}, the resource state is kept, whereas in our case the resource state is traced out after reordering the state $\ket{i_1}\ket{i_2}\dots\ket{i_k}$. 
In the state preparation, the first $k$ qubits being in state $\ket{0^k}$ indicates that there is no repetition among $i_1,i_2,\dots, i_k$.

\begin{lemma}[State preparation for time-ordered integral]\label{lem: prep_integral_state}
For any positive integer $k\geq 2$, there exists a prepare oracle denoted by $\mathrm{PREP}_k^{t}$ such that

\begin{align}
\begin{split}
      \mathrm{PREP}_k^{t} \ket{0^k} \ket{0^{k\log(M)}} \ket{0^{k\log(k)}} \xrightarrow[]{}\sqrt{\frac{k!}{M^k}}\sum_{i_1>i_2>\dots >i_k} \ket{0^k}\ket{i_1}\ket{i_2}\dots \ket{i_k}\sum_{\pi \in S_{k}} \frac{1}{\sqrt{k!}}\ket{\pi} + \ket{\Psi'}
\end{split}
\end{align}
where $\ket{ \Psi'}$ is a unnormalized state, orthogonal to $\ket{0^{k}}\ket{\phi}$ for any $(k\log(M)+k\log(k))$-qubit state $\phi$. 
The oracle uses $\Or(k\log(k)\log(M))$ two-qubit gates and $\Or(k\log(k))$ ancilla qubits.
\end{lemma}

\begin{proof}
The state preparation begins with the state $\ket{0^{k}}\ket{0^{k\log(M)}}$. 
For simplicity, we assume that $M=2^{a}$ for some integer $a$. 
Applying $k\log(M)$ Hadamard gates to the second quantum register yields the state 
\begin{equation}    \sqrt{\frac{1}{M^k}}\sum_{\substack{i_1',i_2',\dots,i_k'=0}}^{M-1}\ket{0^{k}} \ket{i_1'}\dots\ket{i_k'}.
\end{equation}

We use an additional quantum register of size $k\log(k)$ to record the comparison results between $i'_{\ell}$ and $i'_{m}$, for $\ell, m\in\{1,2,\cdots,k\}$ (See \cite{berry2018improved} and classical sorting algorithms for more details). 
Note that we use the comparator of~\cite{berry2018improved}, which outputs $\ket{0}$ if $i'_\ell>i'_{m}$ and $\ket{1}$ otherwise. 
Moreover, if the output is $\ket{1}$, we swap $\ket{i'_\ell}$ and $\ket{i'_m}$. 

For the reordered sequence, we use the first quantum register to post-select the state with no repeated entries within $\ket{i_1'}$, $\ket{i_2'},$\dots, $\ket{i_k'}$. 
For such states, the record of comparisons has a one-to-one correspondence with a permutation $\pi \in S_{p}$. 
We denote by $\ket{\pi}$ the states recording the comparison outcomes, and obtain the state
\begin{equation}
\label{equ:permutation_key}
    \sqrt{\frac{1}{M^{k}}} \sum_{\pi(i_1')>\pi(i_2')>\dots >\pi(i_k')} \ket{0^k} \ket{\pi(i_1')} \ket{\pi(i_2')} \dots \ket{\pi(i_{k}')}\ket{\pi}+\ket{\Psi'}.
\end{equation}
Here $\ket{ \Psi'}$ is a unnormalized state, orthogonal to $\ket{0^{k}}\ket{\psi}$ for any $(k\log(M)+k\log(k))$-qubit state $\phi$. 
Now, we relabel $\pi(i_j')$ by $i_j$ for each $j$. For each state 
\begin{equation}
    \ket{i_1}\dots \ket{i_k},
\end{equation}
there are $k!$ possible permutations. 
Then \cref{equ:permutation_key}
can be rewritten as
\begin{equation}
\label{equ:permutation_key2}
    \sqrt{\frac{1}{M^{k}}} \sum_{\pi \in S_{k}} \sum_{i_1>i_2>\dots >i_k} \ket{0^k} \ket{i_1} \ket{i_2}\dots \ket{i_k}\ket{\pi}+\ket{\Psi'}.
\end{equation}
This can be further simplified as
\begin{equation}
\label{equ:permutation_key3}
    \sqrt{\frac{k!}{M^{k}}}  \sum_{i_1>i_2>\dots >i_k} \ket{0^k} \ket{i_1} \ket{i_2}\dots \ket{i_k} \sum_{\pi \in S_{k}}\sqrt{\frac{1}{k!}}\ket{\pi}+\ket{\Psi'}.
\end{equation}

Finally, we estimate the cost for the state preparation. 
In total, we used $k\log(M)$ Hadamard gates and $\Or(k\log(k))$ comparators. 
Each comparator requires $\Or(\log(M))$ two-qubit gates, so the total cost is $\log(k\log(k)\log(M))$ two-qubit gates.
\end{proof}

\begin{rem}
We can measure the quantum register that records $\pi$ at the end of the algorithm to obtain a pure state

\begin{equation}
\label{equ:permutation_key_4}
    \mathrm{PREP}_k^{t} \ket{0^k} \ket{0^{k\log(M)}} \xrightarrow[]{}\sqrt{\frac{k!}{M^k}}\sum_{i_1>i_2>\dots >i_k} \ket{0^k}\ket{i_1}\ket{i_2}\dots \ket{i_k} + \ket{\Psi}
\end{equation}
due to the structure of $\Psi'$. Here, $\Psi$ is obtained by tracing out the last quantum register, and we use the notation $\Psi$ to emphasize that this state is distinct from $\Psi'$. This prepare oracle provides a more familiar formulation, and together with its adjoint, we can formally construct the block encoding for the Magnus expansion. However, the explicit construction of the block encoding for the Magnus expansion involves the use of additional quantum registers, as shown in the proof, which sets this method apart from the standard block-encoding construction.
\end{rem}

We next discuss the construction of $\text{PREP}^{t,\dagger}_{k}$ and how to apply it using the LCU technique. 
The overall process is essentially the inverse of \cref{lem: prep_integral_state}. Suppose that after applying $\text{PREP}^{t}_{k}$ and a general SELECT oracle we obtain 
\begin{equation}
    \sqrt{\frac{1}{M^{k}}} \sum_{\pi \in S_{k}} \sum_{i_1>i_2>\dots >i_k} \ket{0^k} \ket{i_1} \ket{i_2} \dots \ket{i_k}\ket{\pi} H_{i_1,i_2,\dots, i_k}\ket{\psi}+\ket{\Psi'}\ket{\psi},
\end{equation}
where $H_{i_1,\dots,i_k}$ is a unitary operator and $\psi$ is an $n$-qubit state. 
Based on the comparator record, we can permute $i_j$ back to $i_{j}'=\pi^{-1}(i_j)$ in \cref{lem: prep_integral_state} and obtain
\begin{equation}
    \sqrt{\frac{1}{M^{k}}} \sum_{\pi \in S_{k}} \sum_{i_1>i_2>\dots >i_k} \ket{0^k} \ket{i_1'} \ket{i_2'} \dots \ket{i_k'}\ket{\pi} H_{i_1,i_2,\dots, i_k}\ket{\psi}+\ket{\Psi'}\ket{\psi}.
\end{equation}
Using another $k\log(k)$ comparators on each state $\ket{i_1'}\dots \ket{i_k'}$, we uncompute the $\ket{\pi}$ register and obtain
\begin{equation}
    \sqrt{\frac{1}{M^{k}}} \sum_{\pi \in S_{k}} \sum_{i_1>i_2>\dots >i_k} \ket{0^k} \ket{i_1'} \ket{i_2'} \dots \ket{i_k'}\ket{0^{k\log(k)}} H_{i_1,i_2,\dots, i_k}\ket{\psi}+\ket{\Psi'}\ket{\psi}.
\end{equation}
In the end, we apply $\log(M)$ Hadamard gates to each $\ket{i'}$ register. After post-selecting the first $k+k\log(M)$ qubits, we obtain 
\begin{equation}
\begin{split}
    &\frac{1}{M^{k}} \sum_{\pi \in S_{k}} \sum_{i_1>i_2>\dots >i_k} \ket{0^k} \ket{0^{k\log(M)}}\ket{0^{k\log(k)}} H_{i_1,i_2,\dots, i_k}\ket{\psi}\\
    =&\frac{k!}{M^{k}} \sum_{i_1>\dots>i_k} \ket{0^{k}}\ket{0^{k\log(M)}}\ket{0^{k\log(k)}} H_{i_1,i_2,\dots,i_k}\ket{\psi}.
\end{split}
\end{equation}
It can be observed that the total cost of $\text{PREP}^{t,\dagger}_{k}$ is the same as $\text{PREP}^{t}_{k}$.

\subsubsection{State preparation for Magnus expansion}

State preparation is a crucial step in quantum simulation. 
For \emph{time-independent} quantum simulation, it is embedded in the construction of the input model, where the goal is to prepare the coefficients of the Hamiltonian~\cite{babbush2018encoding}. 
For \emph{time-dependent} quantum simulation, coefficient preparation is also required in constructing the approximated evolution operator. 
In particular, in the $k$-th order Magnus expansion, there are $\mathcal{O}(k \cdot k!)$ coefficients to prepare, which can lead to factorial scaling gate cost.

The prohibitive complexity has hindered further exploration of high-order Magnus expansion-based quantum algorithms. 
However, we propose an approach to construct a prepare oracle, denoted as PREP, for the Magnus expansion that achieves an exponential speedup. 
The key idea is to compute the required coefficients directly on the quantum computer, rather than loading them as classical data. 
The construction of the \text{PREP} is detailed in~\cref{lem: prep_sym_state}.

We apply the direct sampling method introduced in \cite{liu2025block} to transform the binary representation of coefficients into amplitudes on a quantum register. Consider preparing a coefficient $r_b\in [0,1]$ with a binary representation $b$ of length $n_b$. Then the sampling oracle satisfies that
\begin{equation}
	 \text{Sampling}\ \ket{b}\ket{0^{n_b-1}} \ket{0} \to \ket{b}\ket{0^{n_b-1}} \left( r_b \ket{0} + \sqrt{1-r_b^2} \ket{1} \right)+*
\end{equation} 
where $*$ is used to indicate the dropped states after post-selection on the state $\ket{0^{n_b}}$. We note that the quantum register storing 
$b$ is uncomputed in the prepare oracle by reloading the data twice.

\begin{lemma}[State preparation for $\tilde\Omega_k$]\label{lem: prep_sym_state}
 For a positive integer $k\geq 2$, there exists a prepare oracle denoted by $\mathrm{PREP}_k$ 
such that it prepares a superposition of permutations through post-selection, and a sampling oracle prepares the coefficient corresponding to each permutation. To be specific, the prepare oracle and sampling oracle satisfy 
\begin{align}
    &\left \|\bra{0^{n_b}}\otimes \mathbb{I}_{k\log(k)}\left( \mathrm{Sampling } \circ\mathrm{PREP}_k \ket{0^{n_b}} \ket{0^{k\log(k)}} -\frac{\beta_k}{\sqrt{k!}}\sum_{\pi\in S_k} C_{\pi,k}\ket{0^{n_b}}\ket{\pi(0)}\ket{\pi(1)}\dots \ket{\pi(k-1)} \right) \right \| \notag \\
 \leq& \sqrt{C_{\diamond}}k^{\frac{-a+2}{2}} + C_{\diamond}k^{-a+2}=:\tilde{\epsilon}_k ,
\end{align}
where $C_{\diamond}$ is a constant independent of $a$ and $k$, using $\Or(k^2\log^2(k))$ two-qubit gates, $\Or(ak\log (k))$ ancilla qubits  for all constant $a\geq 2$ and satisfies that $C_{\diamond}k^{-a+2}<\frac{3}{4}$. Moreover, the $n_b=\mathcal{O}(\log(k!))$ indicates the number of ancilla qubits used for post-selection and $\frac{1}{2}\leq \beta_k \leq 1$.
\end{lemma}
\begin{proof}
Preparation of coefficients for Magnus expansion can be constructed in two sequential steps:  first, prepare a superposition of permutation states using $\mathrm{PREP}_k$; second, prepare the amplitude $C_{\pi,k}$ through direct sampling and an arithmetic circuit. 
To start, a superposition of permutations can be prepared with methods in \cite{berry2018improved,LiuChildsGottesman2025}

\begin{equation}
    \sqrt{\frac{1-C_{\diamond}k^{-a+2}}{k!}}\sum_{\pi\in S_k}\ket{\pi(0)}\ket{\pi(1)}\dots \ket{\pi(k-1)}+*.
\end{equation}
Here we use $*$ to denote the unnecessary states and their norm can be bounded by $C_{\diamond}k^{-a+2}$. The first step requires $O(ak\log(k))$ ancilla qubits and $O(k\log^2(k))$ two-qubit gates. 

Next, we apply an arithmetic circuit to compute the binary representation of each coefficient $C_{\pi, k}$ based on $\ket{\pi(0)}\ket{\pi(1)}\dots \ket{\pi(k-1)}$. 
Recall that
\begin{equation}
     C _{\pi,k}= \frac{(-1)^{d_a(\pi)}}{k} \times \frac{d_a(\pi)!(k-1-d_a(\pi))!}{ (k-1)!},
\end{equation}
in order to prepare the $C_{\pi,k}$, we prepare the phase $(-1)^{d_a(\pi)}$ and the binary representation of $d_a(\pi)! (k-1-d_a(\pi))!$ separately. With $\mathcal{O}(k)$ times usage of compare oracle, we can compute the number $d_a(\pi)$ through arithmetic addition and prepare the phase $(-1)^{d_a(\pi)}$through Pauli Z gate. Through $k$ arithmetic multiplications on a quantum computer, we can prepare the binary representation of $d_a(\pi)! (k-1-d_a(\pi))!$. The process requires $\mathcal{O}(k^2\log^2(k))$ two qubit gates. In the end, we apply the direct sampling~\cite{liu2025block} to prepare amplitude $C_{\pi,k}$ and uncompute unnecessary quantum registers through $\mathcal{O}(k^2 \log^2(k))$ two-qubit gate. The final state is 

\begin{equation}
    \sqrt{\frac{1-C_{\diamond}k^{-a+2}}{k!}}\sum_{\pi\in S_p} \beta_k C_{\pi,k}\ket{0^{n_b}}\ket{\pi(0)}\ket{\pi(1)}\dots \ket{\pi(k-1)}+*,
\end{equation}
where $n_b = \lceil\log(k!)\rceil$ and $\beta_k = 2^{-n_b+\log(k!)} \in [\frac{1}{2},1]$. The estimate follows naturally from $C_{\diamond}k^{-a+2}<\frac{3}{4}$.

Combining the resource costs of both oracles gives the overall resource scaling. 
\end{proof}

The construction of a superposition of permutation states requires an additional quantum register, which is not accounted for in the proof. This distinction is similar to the difference between \cref{equ:permutation_key3} and \cref{equ:permutation_key_4}. The procedure for constructing $\text{PREP}^{\dagger}_{k}$ follows an approach analogous to the time-ordered integral case. For brevity, we omit the detailed discussion here.

\subsubsection{Third order Magnus expansion}

\begin{figure}[htbp!]
\centering
\begin{tikzpicture}
\node[scale=0.725]{
   \begin{quantikz}[column sep=0.325cm]
\lstick{$\ket{0}^{\otimes 3}$} &\qwbundle{}&&\gate[5]{\shortstack{PREP}_3^t}&&\gategroup[13,steps=9,style={dashed,rounded corners},background,label style={label position=below,anchor=north,yshift=-0.2cm}]{SELECT}&&&&&&&&&\gate[5]{\shortstack{PREP}_3^{t\dagger}}&&\meter[12]{}\\
  \lstick{$\ket{0}^{\otimes \log (M)}$}&\qwbundle{}\wire[l][1]["p"{above,pos=0.5}]{a}&&&&\gate[3]{\shortstack{SWAP\\UP}}&\ctrl[style=U]{10}&\gate[3]{\shortstack{SWAP\\UP}}&\gate[3]{\shortstack{SWAP\\UP}}&\ctrl[style=U]{10}&\gate[3]{\shortstack{SWAP\\UP}}&\gate[3]{\shortstack{SWAP\\UP}}&\ctrl[style=U]{10}&\gate[3]{\shortstack{SWAP\\UP}}&&&\\
  \lstick{$\ket{0}^{\otimes \log (M)}$}&\qwbundle{}\wire[l][1]["q"{above,pos=0.5}]{a}&&&&&&&&&&&&&&&\\
  \lstick{$\ket{0}^{\otimes \log (M)}$}&\qwbundle{}\wire[l][1]["r"{above,pos=0.5}]{a}&&&&&&&&&&&&&&&\\
  \lstick{$\ket{0}^{\otimes 3\lceil\log (3)\rceil}$} &\qwbundle{}&&&&&&&&&&&&&&&\\
\lstick{$\ket{0}^{\otimes n_b}$} & &&\gate[4]{\shortstack{PREP}_3}&\gate[4]{\shortstack{Sampling}}&&&&&&&&&&\gate[4]{\shortstack{PREP}_3^\dagger}&&\\
  \lstick{$\ket{0}^{\otimes \lceil\log (3)\rceil}$}&\qwbundle{}&&&&\ctrl[style=Uf]{-3}&&\ctrl[style=Uf]{-3}&&&&&&&&&\\
  \lstick{$\ket{0}^{\otimes \lceil\log (3)\rceil}$}&\qwbundle{}&&&&&&&\ctrl[style=Uf]{-4}&&\ctrl[style=Uf]{-4}&&&&&&\\
  \lstick{$\ket{0}^{\otimes \lceil\log (3)\rceil}$}&\qwbundle{}&&&&&&&&&&\ctrl[style=Uf]{-5}&&\ctrl[style=Uf]{-5}&&&\\
  \lstick{$\ket{0}^{\otimes n_a}$}&\qwbundle{}&&&&&&&&&\swap{2}&&&&&&\\
  \lstick{$\ket{0}^{\otimes n_a}$}&\qwbundle{}&&&&&&\swap{1}&&&&&&&&&\\
  \lstick{$\ket{0}^{\otimes n_a}$}&\qwbundle{}&&&&&\gate[2]{\shortstack{Ham\\-T}}&\targX{}&&\gate[2]{\shortstack{Ham\\-T}}&\targX{}&&\gate[2]{\shortstack{Ham\\-T}}&&&&\\
  \lstick{$\ket{\psi}$}&\qwbundle{}&&&&&&&&&&&&&&&
\end{quantikz}
};
\end{tikzpicture}
    \caption{ The quantum circuit for the oracle $O_{\tilde{\Omega}_3}$ that block encodes $\Tilde{\Omega}_3(t_j+h,t_j)$ for $t_j=0$ in \cref{eqn: tilde_omega_3}. 
    }
    \label{fig:omega_3_circuit}
\end{figure}
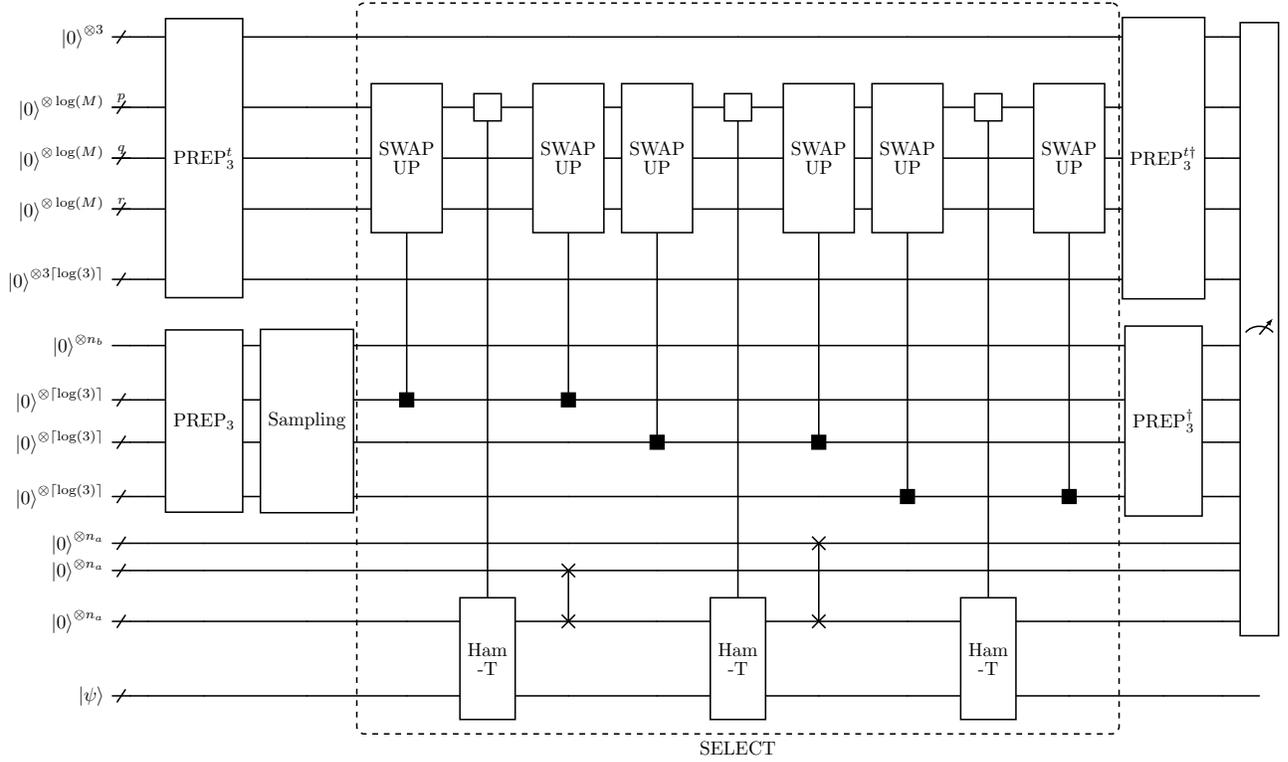

We now construct the quantum circuit for the Magnus expansion of arbitrary order. To block encode the $p$-th order Magnus expansion, we construct a quantum circuit to block encode each $\Omega_k(t_j+h,t_j)$ term in $\Omega_{(p)}(t_j+h,t_j)$ for $1\leq k\leq p$. Then, with the LCU technique, we combine them to obtain the input model for the $p$-th order Magnus expansion. 

As an example, consider the $\Omega_{(3)}(t_j+h,t_j)$ and we use $M$ time steps to discretize the multi-layer time integral from $t_j$ to $t_j+h$. By~\cref{eqn:omega_k_expansion} the highest order term $\Omega_3(t_j+h,t_j)$ in the third-order Magnus expansion  $\Omega_{(3)}(t_j+h,t_j)$ equals
\begin{equation}
    \Omega_3(t_j+h,t_j) = \sum_{\pi\in S_3} C_{\pi,3} \int_{t_j}^{t_j+h} d\tilde{t}_1  \int_{t_j}^{\tilde{t}_1} d\tilde{t}_2 \int_{t_j}^{\tilde{t}_{2}} d\tilde{t}_{3}\, A(\tilde{t}_{\pi(1)})A(\tilde{t}_{\pi(2)})A(\tilde{t}_{\pi(3)}).
\end{equation}
For a total discretization step count $M$ , the discrete implementation of $\Omega_3(t_j+h,t_j)$ equals
\begin{equation}\label{eqn: tilde_omega_3}
\begin{split}
    \Tilde{\Omega}_3(t_j+h,t_j) &= \sum_{i_1=0}^{M-1} \sum_{i_2=0}^{i_1-1} \sum_{i_3=0}^{i_2-1}\sum_{\pi\in S_3} C_{\pi,3}\, A\left(t_j+\frac{ i_{\pi(1)}h}{M} \right)A\left( t_j+\frac{i_{\pi(2)}h}{M} \right) \\
    &\times A\left( t_j + \frac{i_{\pi(3)}h}{M}\right) \frac{h^3}{M^3}.
\end{split}
\end{equation}

In the construction of quantum circuit for $\Tilde{\Omega}_3(t_j+h,t_j)$, a naive implementation of SELECT oracle, 
\begin{equation}
\begin{split}
    &\text{SELECT} \ \ket{i_1}\ket{i_2}\ket{i_3}\ket{\pi(1)} \ket{\pi(2)} \ket{\pi(3)} \ket{0^{3n_a}}\ket{\psi}\\ &\xrightarrow[]{} \ket{i_1}\ket{i_2}\ket{i_3}\ket{\pi(1)} \ket{\pi(2)} \ket{\pi(3)} \ket{0^{3n_a}}\frac{A\left(t_j+\frac{ i_{\pi(1)}h}{M} \right)}{\alpha} \frac{A\left(t_j+\frac{ i_{\pi(2)}h}{M} \right)}{\alpha} 
    \frac{A\left(t_j+\frac{ i_{\pi(3)}h}{M} \right)}{\alpha}\ket{\psi}+* \,,
\end{split}
\end{equation}
would require at least $\mathcal{O}(3!)$ multi-controlled gates. In general, such construction requires $\mathcal{O}(p!)$ multi-controlled gates for $\tilde{\Omega}_p(t_j+h,t_j)$.
However, by employing the SWAP-UP circuit introduced in~\cite{low2024trading}, the gate cost of constructing the SELECT oracle can be reduced to $\mathcal{O}(p)$~\cite{Wan2021exponentially,liu2025block}, which is negligible compared with the overall cost of block encoding. The SWAP-UP is defined by
\begin{equation}
    \mbox{SWAP-UP} : \ket{p}\ket{j_0}\ket{j_1} \cdots \ket{j_{n-1}} \to \ket{p}  \ket{j_p} \ket{*}\dots\ket{*}\,,
    \label{eq:swapup}
\end{equation}
where $j_i\in \{0,1\}$ for $0\leq i \leq n-1$, $0\leq p \leq n-1$ and $n$ is a positive integer. 
We adopt a generalized formulation in which $j_i$ is the $\log(M)$-bit binary representation of the index $i_1$, $i_2$, or $i_3$. 
The detailed construction of the SELECT oracle for the Magnus expansion is presented in \cref{lem:3order} and illustrated in \cref{fig:omega_3_circuit}.
\begin{lemma}
\label{lem:3order}
The quantum circuit in \cref{fig:omega_3_circuit} is a $(\alpha^3h^3/\beta_3 , 3n_m+n_b+6\lceil \log(3)\rceil +3n_a+3,\tilde{\epsilon}_3)$-block encoding of $\Tilde{\Omega}_3$ where $\tilde \epsilon_3:=\sqrt{C_{\diamond}}3^{\frac{-a+2}{2}} + C_{\diamond}3^{-a+2}$ where $\beta_3 = 2^{-\lceil \log(3!)\rceil+\log(3!)} \in [\frac{1}{2},1]$ . 
\end{lemma}
\begin{proof}
To start, the initial state is set to be
\begin{equation}
     \ket{0^3}\ket{0^{3 \log(M)}}\ket{0^{3\lceil \log(3) \rceil}}   \ket{0^{n_b}} \ket{0^{\lceil \log(3) \rceil}} \ket{0^{\lceil \log(3) \rceil}} \ket{0^{\lceil\log(3)\rceil}}\ket{0^{3n_a}} \ket{\psi}
\end{equation}
where $\ket{\psi}$ is a $n$-qubit state. With the $\text{PREP}_{3}^{t}$ in \cref{lem: prep_integral_state}, we have 

\begin{equation}
    \sqrt{\frac{3!}{M^{3}}} \sum_{i_1>i_2>i_3} \ket{0^3}\ket{i_1}\ket{i_2} \ket{i_3} \left(\sum_{\hat{\pi} \in S_{3}}\sqrt{\frac{1}{3!}}\ket{\hat{\pi}} \right)  \ket{0^{n_b}} \ket{0^{\lceil \log(3) \rceil}} \ket{0^{\lceil \log(3) \rceil}}\ket{0^{\lceil \log(3) \rceil}} \ket{0^{3n_a}} \ket{\psi} + *\,.
\end{equation}
After that, we apply the $\text{PREP}_3$ and Sampling oracle %
to prepare the coefficients within the Magnus expansion and have 
\begin{equation}
    \sqrt{\frac{3!}{M^{3}}} \sum_{i_1>i_2>i_3} \frac{\beta_3}{\sqrt{3!}}\sum_{\pi \in S_{3}} C_{\pi,3}\ket{0^3}\ket{i_1}\ket{i_2} \ket{i_3}\left(\sum_{\hat{\pi} \in S_{3}}\sqrt{\frac{1}{3!}}\ket{\hat{\pi}} \right)  \ket{0^{n_b}} \ket{\pi(0)} \ket{\pi(1)}\ket{\pi(2)} \ket{0^{3n_a}} \ket{\psi}+*\,.
\end{equation}

In the next step, by applying HAM-T oracles through the SWAP-UP  gates (SELECT in \cref{fig:omega_3_circuit}) in \cite{Wan2021exponentially,low2024trading}, we obtain
\begin{equation}
    \begin{split}
        &\sqrt{\frac{3!}{M^{3}}} \sum_{i_1>i_2>i_3} \frac{\beta_3}{\alpha^3\sqrt{3!}}\sum_{\pi \in S_{3}} C_{\pi,3}\ket{0^3}\ket{i_1}\ket{i_2} \ket{i_3} \left(\sum_{\hat{\pi} \in S_{3}}\sqrt{\frac{1}{3!}}\ket{\hat{\pi}} \right) \ket{0^{n_b}} \ket{\pi(0)} \ket{\pi(1)}\ket{\pi(2)} \ket{0^{3n_a}}\\
        &\times A\left(t_j+\frac{ i_{\pi(1)}h}{M} \right)A\left( t_j+\frac{i_{\pi(2)}h}{M} \right)A\left( t_j + \frac{i_{\pi(3)}h}{M}\right)\ket{\psi}+* \,,
    \end{split}
\end{equation}
where the factor $1/\alpha^3$ comes from three applications of the HAM-T oracles.
Finally, applying $\text{PREP}_{3}^{\dagger}$ and $\text{PREP}_{3}^{t,\dagger}$ yields
\begin{equation}
    \begin{split}
        &  \frac{\beta_3}{M^{3}\alpha^3} \sum_{i_1>i_2>i_3} \sum_{\pi \in S_3}C_{\pi,3}\ket{0^3}\ket{0^{\lceil \log(3) \rceil}}\ket{0^{\lceil \log(3) \rceil}} \ket{0^{\lceil \log(3) \rceil}} \ket{0^{3\lceil \log(3) \rceil}} \ket{0^{n_b}} \ket{\pi(0)} \ket{\pi(1)}\ket{\pi(2)} \ket{0^{3n_a}}\\
        &\times A\left(t_j+\frac{ i_{\pi(1)}h}{M} \right)A\left( t_j+\frac{i_{\pi(2)}h}{M} \right)A\left( t_j + \frac{i_{\pi(3)}h}{M}\right)\ket{\psi}+*.
    \end{split}
\end{equation}

\end{proof}

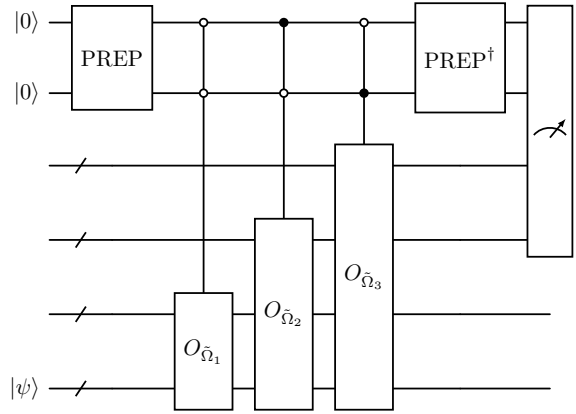
\begin{figure*}[htbp!]
\centering
\begin{tikzpicture}
\node[scale=0.85]{
   \begin{quantikz}[column sep=0.35cm]
\lstick{$\ket{0}$}&\gate[2]{\text{PREP}}&\ctrl[open]{4}&\ctrl[]{3}&\ctrl[open]{2}&\gate[2]{\text{PREP}^\dagger}&\meter[4]{}\\
\lstick{$\ket{0}$}&&\ctrl[open]{0}&\ctrl[open]{0}&\ctrl[]{0}&&\\
\lstick{}&\qwbundle{}&&&\gate[4]{O_{\tilde\Omega_3}}&&\\
\lstick{}&\qwbundle{}&&\gate[3]{O_{\tilde\Omega_2}}&&&\\
\lstick{}&\qwbundle{}&\gate[2]{O_{\tilde\Omega_1}}&&&&\\
\lstick{$\ket{\psi}$}&\qwbundle{}&&&&&
\end{quantikz}
};
\end{tikzpicture}
    \caption{The quantum circuit of oracle $O_{\tilde{\Omega}_{(3)}}$ that block encodes $\Tilde{\Omega}_{(3)}(t_j+h,t_j)$ for $t_j=0, p=3$ in \cref{eqn: Omega_(p)}.}
    \label{fig:omega_(3)_LCU}
\end{figure*}

 For the second-order terms, we similarly have $(\alpha^2h^2/\beta_2, 2n_m+n_b+4\lceil \log(2)\rceil +2n_a+2,\sqrt{C_{\diamond}}2^{\frac{-a+2}{2}} + C_{\diamond}2^{-a+2})$ block encoding of $\tilde\Omega_2$. For the first-order term, the use of complex prepare oracles is not required. We refer the reader to~\cite{AnFangLin2021,FangLiuSarkar2025} for a detailed construction of the corresponding block encoding. Note that since $|S_1|=1$, the subnormalization can be improved to $\alpha h$. By applying a linear combination of unitaries and a sequence of control rotation gates (used to unify the factor $\beta_k$ to $\frac{1}{2}$), we can show that the circuit in \cref{fig:omega_(3)_LCU} gives a block encoding for $\Omega_{(3)}$.

 \begin{lemma}
 Assume $\alpha h \leq \gamma<1$, then quantum circuit in \cref{fig:omega_(3)_LCU} gives a $$\left(2C^{\gamma}_{(3)}\alpha h,3n_m+n_b+6\lceil \log(3)\rceil +3n_a+3,\tilde{\epsilon}_1+ \tilde{\epsilon}_2+\tilde{\epsilon}_3 \right)$$ block encoding of $\Tilde{\Omega}_{(3)}$ where $$C^{\gamma}_{(3)}=1 + \alpha h+ \alpha^2 h^2.$$
 \end{lemma}

 \begin{proof}
 To start, the initial state is set to be $\ket{0^2}\ket{0^{n_3}}\ket{\psi}$ where $n_3:=3n_m+n_b+3\lceil \log(3)\rceil +3n_a+3$. We constructed the prepare oracle (denoted as \text{PREP}) with alias sampling~\cite{babbush2018encoding}, 

 \begin{equation}
     \text{PREP} \ket{0}\ket{0} = \sqrt{\frac{\alpha h}{\sum_{i=1}^{3} \alpha^i h^i} }\ket{1} +  \sqrt{\frac{\alpha^2 h^2}{\sum_{i=1}^{3} \alpha^i h^i} }\ket{2} + \sqrt{\frac{\alpha^3 h^3}{\sum_{i=1}^{3} \alpha^i h^i} } \ket{3}.  
 \end{equation}
Then we have that
\begin{equation}
   \ket{0^2}\ket{0^{n_3}}\ket{\psi} \xrightarrow[]{\text{PREP}} \left( \sqrt{\frac{\alpha h}{\sum_{i=1}^{3} \alpha^i h^i}} \ket{1} +  \sqrt{\frac{\alpha^2 h^2}{\sum_{i=1}^{3} \alpha^i h^i} }\ket{2} + \sqrt{\frac{\alpha^3 h^3}{\sum_{i=1}^{3} \alpha^i h^i}} \ket{3} \right) \ket{0^{n_3}} \ket{\psi}.
\end{equation}
With control on the top two qubits for $\ket{1}$, $\ket{2}$ and $\ket{3}$, the block encoding of Magnus operators act on the state $\ket{\psi}$. That gives
\begin{equation}
     \sqrt{\frac{\alpha h}{\sum_{i=1}^{3} \alpha^i h^i}} \ket{1}\ket{0^{n_3}} \frac{\tilde{\Omega}_1}{2\alpha h }\ket{\psi} +  \sqrt{\frac{\alpha^2 h^2}{\sum_{i=1}^{3} \alpha^i h^i} }\ket{2}\ket{0^{n_3}} \frac{\tilde{\Omega}_2}{2\alpha^2 h^2 }\ket{\psi}  + \sqrt{\frac{\alpha^3 h^3}{\sum_{i=1}^{3} \alpha^i h^i}} \ket{3}  \ket{0^{n_3}} \frac{\tilde{\Omega}_3}{2\alpha^3 h^3 }\ket{\psi} + *
\end{equation}
 where $*$ refers to the dropped states after post-selection in the end. In the end, the $\text{PREP}^{\dagger}$ gives
 \begin{equation}
 \begin{split}
     \frac{1}{2\sum_{i=1}^{3} \alpha^i h^i} \sum_{k=1}^{3} \ket{0^2} \ket{0^{n_3}} \tilde{\Omega}_{k}  \ket{\psi}
     = \frac{1}{2\sum_{i=1}^{3} \alpha^i h^i}  \ket{0^2} \ket{0^{n_3}} \tilde{\Omega}_{(3)}  \ket{\psi}.
 \end{split}
 \end{equation}
 In the end, we define subnormalization factor $2C^{\gamma}_{(3)}$ where $C^{\gamma}_{(3)}=1 + \alpha h+ \alpha^2 h^2$. With $\alpha h \leq \gamma <1$, the subnormalization is of order $\mathcal{O}(1)$. We note that each $\tilde{\Omega}_k$ has error $\tilde{\epsilon}_k$ and the error propagate to the full block encoding linearly.
 \end{proof}

\subsubsection{High order Magnus expansion}

Recall that the high order Magnus expansion $\Tilde{\Omega}_{(p)}(t_j+h,t_j)$ for a short time evolution, satisfies that
\begin{equation}\label{eqn: Omega_(p)}
    \Tilde{\Omega}_{(p)}(t_j+h,t_j) =\sum_{k=1}^{p} \tilde\Omega_k (t_j+h,t_j)
\end{equation}
where
\begin{equation}
    \begin{split}
&\Omega_k(t_j+h,t_j) = \sum_{\pi\in S_k} C_{\pi,k} \int_{t_j}^{t_j+h} d\tilde{t}_1 \int_{t_j}^{\tilde{t}_1} d\tilde{t}_2  \dots  \int_{t_j}^{\tilde{t}_{k-1}} d\tilde{t}_{k} A(\tilde{t}_{\pi(1)})A(\tilde{t}_{\pi(2)})\dots A(\tilde{t}_{\pi(k)}).
    \end{split}
\end{equation}

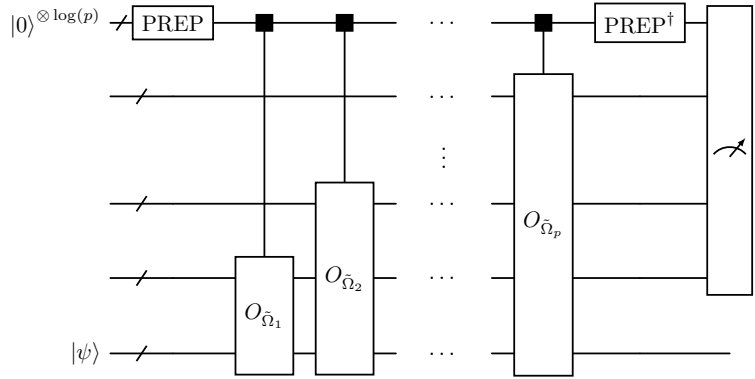
\begin{figure*}[htbp!]
\centering
\begin{tikzpicture}
\node[scale=0.85]{
\begin{quantikz}[column sep=0.35cm]
\lstick{$\ket{0}^{\otimes\log (p)}$}&\gate{\text{PREP}}\qwbundle{}&\ctrl[style=Uf]{4}&\ctrl[style=Uf]{3}&\push{\makebox[1.5cm]\dots}&\ctrl[style=Uf]{1}&\gate{\text{PREP}^\dagger}&\meter[5]{}\\
\lstick{}&\qwbundle{}&&&\push{\makebox[1.5cm]\dots}&\gate[5]{O_{\tilde\Omega_p}}&&\\
\setwiretype{n} \lstick{}  &&&&\midstick[brackets=none]{\vdots}&&&\\
\lstick{}&\qwbundle{}&&\gate[3]{O_{\tilde\Omega_2}}&\push{\makebox[1.5cm]\dots}&&&\\
\lstick{}&\qwbundle{}&\gate[2]{O_{\tilde\Omega_1}}&&\push{\makebox[1.5cm]\dots}&&&\\
\lstick{$\ket{\psi}$}&\qwbundle{}&&&\push{\makebox[1.5cm]\dots}&&&
\end{quantikz}
};
\end{tikzpicture}
    \caption{The quantum circuit of oracle $O_{\tilde{\Omega}_{(p)}}$ that block encodes $\Tilde{\Omega}_{(p)}(t_j+h,t_j)$ for $t_j=0$ in \cref{eqn: Omega_(p)}.}
    \label{fig:omega_(p)_LCU}
\end{figure*}

The circuit design in \cref{fig:omega_3_circuit} can be extended for any order Magnus term $\tilde \Omega_k$ in the Magnus expansion $\tilde \Omega_{(p)}$ for $1\leq k\leq p$. Using the LCU method with $\text{PREP}$ and $\text{PREP}^{\dagger}$ as shown in \cref{fig:omega_(3)_LCU}, we can extend the quantum circuit of the Magnus expansion of order $3$ to any order $p$. 
\cref{fig:omega_(p)_LCU} shows the construction of quantum circuit of block encoding for $p$-th order Magnus expansion $\Tilde{\Omega}_{(p)}$.

State preparation can be a crucial task for implementing high-order Magnus expansion, in order to avoid factorial or exponential cost in two-qubit gate count. 
To construct the block encoding for high-order Magnus expansion with $\Or(\epsilon)$ error, a necessary condition is that the error from $\text{PREP}$ and Sampling oracle is no more than $\epsilon$. 
We summarize the analysis of gate complexity and total number of ancilla qubits for the state preparation for arbitrary $p$-th order Magnus expansion in \cref{lem: prep_state_Omega(p)}.

\begin{lemma}[State preparation for $\tilde\Omega_{(p)}$]\label{lem: prep_state_Omega(p)}
    For a positive integer $p$, preparing the states for \ $\Omega_k, k\in\{1,2,\cdots,p\}$ in $\Omega_{(p)}$ requires $\Or(p\log(p)\log(M)+p^3\log^2(p))$ two-qubit gates, and $\Or(p^2\log (p) \log(\frac{p}{\epsilon}))$ ancilla qubits, for a target error $\epsilon$.
\end{lemma}
\begin{proof}

    By~\cref{lem: prep_sym_state}, the total error is $\sum_{k=1}^p\tilde\epsilon_k$ where $\tilde\epsilon_k$ indicates the state preparation error for the $k$-th order Magnus expansion term and we recall its expression as
    \begin{equation}
      \tilde{\epsilon}_k =   \sqrt{C_{\diamond}}k^{\frac{-a+2}{2}} + C_{\diamond}k^{-a+2}. 
    \end{equation}
    Note that one can also tune the choice of $a$ according to $k$ (denoted as $a_k$). To be specific, it is sufficient to require $\frac{1}{\sqrt{k^{a_{k}-2}}}\leq \frac{\epsilon}{2\sqrt{C_{\diamond}}p}$ for all $k>1$. Therefore, we can set $a_k=\mathcal{O}(\log_{k}(\frac{p}{\epsilon}))$. With this choice of $a_k$ for the target error $\epsilon$, we then propagate $a_k$ to the gate and ancilla count. 
     The state preparation for the Magnus expansion requires
    \begin{equation}
        p+\sum_{k=2}^{p} k^2\log^{2}(k) =\mathcal{O}(p^3\log^2(p)),
    \end{equation}
    two-qubit gate and the state preparation for the time-ordered integral requires
    \begin{equation}
        \sum_{k=2}^{p} k \log(k) \log(M)  =\mathcal{O}(p\log(p)\log(M))
    \end{equation}
    two-qubit gates. In total, we require

\begin{equation}
    \mathcal{O}(p^3\log^2(p)+p\log(p)\log(M))
\end{equation}
two-qubit gates.     
   Also, the number of ancilla qubits used is 
    \begin{equation}
         \sum_{k=2}^{p} a_k k \log(k) +  k\log(k) = \mathcal{O} \left(p^2\log(p) \log\left(\frac{p}{\epsilon}\right) \right).
    \end{equation}
\end{proof}

This section concludes with a summary of the results pertaining to the complexity analysis of block encoding for high-order Magnus expansions.

\begin{lemma}\label{lemma: pth_order_omega}
 Assume $\alpha h \leq \gamma<1$, the quantum circuit in \cref{fig:omega_(p)_LCU} gives a $(2C_{(p)}^{\gamma}\alpha h,p\log(M)+n_b+2p\lceil \log(p)\rceil +pn_a+p,\epsilon)$ block encoding of $\Tilde{\Omega}_{(p)}$ where
 \begin{equation}
     1 \leq C^{\gamma}_{(p)} < \frac{1}{1-\gamma},
 \end{equation}
 and the quantum circuit requires $\mathcal{O}(p^3\log^2(p)+p\log(p)\log(M))$ two-qubit gate and $\mathcal{O}(p^2\log(p)\log(\frac{p}{\epsilon})+p n_a +p\log(M))$ ancilla qubits.
 \end{lemma}

\section{Long-time complexity of high-order Magnus expansion}\label{sec: cost}
In this section, we analyze the total resource costs required to simulate quantum dynamics using high-order Magnus expansion on a quantum computer. We first consider the case where the Magnus expansion order $p$ is held constant, and derive explicit cost scaling for a target simulation accuracy. We then generalize to the setting where the Magnus order can grow with the desired precision.

We first establish the necessary choice of step size and number of slices to control the global Magnus truncation error.

\begin{lemma}[Parameter selection for global Magnus error with constant order]\label{lem: long_time_magnus_error_constantorder}
Let $p$ be a fixed positive integer, and let $\epsilon > 0$ be the target global error. Define
\begin{equation}\label{eq:def_bar_alpha_comm}
\bar \alpha_\mathrm{comm} := \max_{p+1 \leq q \leq p^2+2p} \alpha_{\mathrm{comm}, q}^{1/q}.
\end{equation}
To ensure that the global Magnus expansion error is bounded by $\epsilon$, it suffices to choose the following parameters
\begin{equation}
 \quad L = \mathcal{O}{ \left( \frac{\bar \alpha_{\mathrm{comm}}^{1+1/p}\, T^{1+1/p}}{ \epsilon^{1/p}} \right)}.
\end{equation}
\end{lemma}
\begin{proof}
By~\cref{thm:magnus_trunction_finite_comm}, the long-time Magnus error can be bounded as 
\begin{align}
    \norm{U(T,0)-U_p(T,0)} 
&\leq \frac{1}{p+1} \bar\alpha_\mathrm{comm}^{p+1}h^p T +C \sum_{q=p+2}^{p^2+2p} \bar\alpha_\mathrm{comm}^{q} h^{q-1}T \notag\\
&= \bar\alpha_\mathrm{comm}^{p+1}h^p T\left(\frac{1}{p+1} + C \sum_{q=p+2}^{p^2+2p}\bar\alpha_\mathrm{comm}^{q-(p+1)}h^{q-(p+1)}\right) = \bar\alpha_\mathrm{comm}^{p+1}h^p T C_1. \tag{for some constant $C_1$}
    \end{align}
Setting the long-time Magnus error equal to $\epsilon$ gives
\begin{align}
     L &= \frac{C_1^{1/p}\,\bar\alpha_\mathrm{comm}^{1+1/p}\,T^{1+1/p}}{\epsilon^{1/p}} = \Or\left(\frac{\bar\alpha_\mathrm{comm}^{1+1/p}\,T^{1+1/p}}{\epsilon^{1/p}} \right).
\end{align}

\end{proof}

Next, we want to ensure that the error from discretizing the time integrals does not exceed the target precision and is matched to the Magnus truncation error itself. This motivates the following result, which bounds the discretization steps.

\begin{lemma}\label{lem: discretization_eqaulto_magnus_constantorder}
    For constant order $p$, to make the long-time Magnus error match the long-time quadrature error, with the parameters fixed in \cref{lem: long_time_magnus_error_constantorder}, it is sufficient to choose the discretization steps $M$ to be
    \begin{equation}
        M = \Or\left( \frac{\|A'\|\,T^{1-1/p}}{\bar \alpha_\mathrm{comm}^{1+\frac{1}{p}}\,\epsilon^{1-1/p}}\right).
    \end{equation}
\end{lemma}
\begin{proof}
    We start from the local quadrature error in \cref{thm: local_quadrature_error},
    \begin{align}
        &\sum_{k=1}^{p}\frac{1}{2M}\left( 2k h^{k+1}  \|A'\| \|A\|^{k-1} + (k-1) h^k \|A\|^k\right)\notag\\
        =& \frac{1}{2M}\left(2h^2 \|A'\|+4h^3\|A'\|\|A\|+h^2\|A\|^2+\cdots + 2ph^{p+1}\|A'\|\|A\|^{p-1}+(p-1)h^p\|A\|^p\right)\label{eqn:expanded_local_quadrature_error}.
    \end{align}
    Collecting all the non $\|A'\|$ terms in the parathesis of \cref{eqn:expanded_local_quadrature_error}, assume $\alpha h<1$,  we have 
    \begin{align*}
        \sum_{k=1}^p (k-1)(h\|A\|)^{k} &\leq \sum_{k=1}^p (k-1)(h\alpha)^k =C,
    \end{align*}
    for some constant $C$. Collecting all the $\|A'\|$ terms in the parathesis of \cref{eqn:expanded_local_quadrature_error} yields:
    \begin{align}
        h^2\|A'\| \left(2+4h\|A\|+6(h\|A\|)^2 +\cdots +2p(h\|A\|)^{p-1}\right) = h^2 \|A'\| C_1,\notag
    \end{align}
    for some constant $C_1$. Therefore, \cref{eqn:expanded_local_quadrature_error} becomes $\frac{C_2 h^2 \|A'\|}{M}$ for some constant $C_2$. Last, setting the long-time Magnus error in \cref{lem: long_time_magnus_error_constantorder} equal to the long-term quadrature error, for some constant $C_3$ we have
    \begin{align*}
        \frac{C_3 L h^2 \|A'\|}{M} & = \bar\alpha_\mathrm{comm}^{p+1}h^p T,
    \end{align*}
thus
\begin{align}
    M&= C_3 \frac{Lh^2\|A'\|}{\bar\alpha_\mathrm{comm}^{p+1}\,h^pT} =C_3 \frac{\| A'\|}{\bar\alpha_\mathrm{comm}^{p+1}\, h^{p-1}}\notag \\
    &= C_3 \| A'\|\bar\alpha_\mathrm{comm}^{-2} \left(\frac{L}{\bar\alpha_\mathrm{comm} T} \right)^{p-1} = \Or\left( \frac{\|A'\|\,T^{1-1/p}}{\bar \alpha_\mathrm{comm}^{1+\frac{1}{p}}}\,\epsilon^{1-1/p}\right),
\end{align}
where the last equality used the $L = \mathcal{O}{ \left( \frac{\bar \alpha_{\mathrm{comm}}^{1+1/p}\, T^{1+1/p}}{ \epsilon^{1/p}} \right)}$.
\end{proof}

By matching the quadrature error to the Magnus truncation error, we guarantee that neither source of error dominates the total simulation accuracy. 
With the choice of parameters $L$ and $M$ fixed as above, we are now ready to analyze the total quantum resource cost required for implementing the constant-order Magnus simulation.

\begin{thm}[Resource cost for $p$-th order Magnus expansion]\label{thm: total_resource_cost_constantorder}
Let the Hamiltonian $H(t)$ satisfies that $\|H(t)\|\leq\alpha$ for all $t$, 
 and $U(T,0) = \mathcal{T}e^{-i\int_0^T H(s) \, ds}$ be its exact long-time propagator over the time interval $[0, T]$. 
 Let $U_p(T,0)$ denote the time evolution of $p$-th order Magnus approximation  as the product of the slice-wise propagators:
\begin{equation}
    U_p(T,0)=\prod_{j=0}^{L-1} U_{p}\left(t_{j+1},t_{j}\right)
\end{equation}
where $t_j = j h$, $h=T/L$.
Then for a given $\epsilon$, $p$, and $T$, there exists a quantum circuit $\mathcal C$ that implements $U_p(T,0)$ with total error $\|\mathcal C - U(T,0)\|\leq4\epsilon$, failure probability $\Or(\epsilon)$, and the following cost:
\begin{enumerate}
    \item $\Or\left(p^2\left(\alpha T+\frac{\bar \alpha_\mathrm{comm}^{1+1/p}\, T^{1+1/p}}{\epsilon^{1/p}}\log\left(\frac{\bar \alpha_\mathrm{comm}\,T}{\epsilon}\right)\right)\right)$  uses of HAM-T oracles,
    \item $\Or\left(p\left(\log\left( \frac{\|A'\|\,T^{1-1/p}}{\bar \alpha_\mathrm{comm}^{1+\frac{1}{p}}\,\epsilon^{1-1/p}}\right)+p\log (p)\log\left(p \frac{\bar\alpha_\mathrm{comm} T}{\epsilon} \right)\right)\right)$ ancilla qubits,
    \item $\Or\left(\left(p^3\log^2(p)+p\log (p)\log\left( \frac{\|A'\|\,T^{1-1/p}}{\bar \alpha_\mathrm{comm}^{1+\frac{1}{p}}\,\epsilon^{1-1/p}}\right)\right)\cdot\left(\alpha T+\frac{\bar \alpha_\mathrm{comm}^{1+1/p}\, T^{1+1/p}}{\epsilon^{1/p}}\log\left(\frac{\bar \alpha_\mathrm{comm}\,T}{\epsilon}\right)\right)\right)$  uses of two-qubit gates. 
\end{enumerate}
\end{thm}
\begin{proof}
We first show the HAM-T count by analyzing the implementation of each $U_p(t_{j+1},t_j)$. 
By \cref{lemma: pth_order_omega}, the $p$-th order discretized Magnus expansion $\tilde \Omega_{(p)}(t_j+h,t_j)$ has a $(C_{(p)}^{\gamma}\alpha h,\lceil \log(|S_p|)\rceil +pn_a+p+2,\epsilon/L)$ block encoding, therefore by~\cite{GilyenSuLowEtAl2019}, its time evolution $\exp\left(\tilde \Omega_{(p)}(t_j+h,t_j)\right)$ has a $(1,\log(S_p)+pn_a+p+4,\delta+\epsilon/L)$ block encoding, and it uses the $\tilde \Omega_{(p)}$ circuit   $C_{(p)}^{\gamma}\alpha h+\log(\frac{1}{\delta})$ times.  
By \cref{fig:omega_3_circuit} and \cref{fig:omega_(p)_LCU}, each $\tilde \Omega_{(p)}$ invokes HAM-T $\sum_{k=1}^{p}k<p^2$ times.
Therefore, the circuit for $\exp\left(\tilde \Omega_{(p)}(t_j+h,t_j)\right)$ invokes HAM-T $p^2\cdot(C_{(p)}^{\gamma}\alpha h+\log(\frac{1}{\delta}))$ times. Consequently, the total number of HAM-T for a circuit implementing $\prod_{j=0}^{L-1}\exp\left(\tilde \Omega_{(p)}(t_j+h,t_j)\right)$ is
\begin{align}\label{eqn: temp_HAMT_cost_constant_p}
    L \cdot p^2 \cdot \left(C_{(p)}^{\gamma}\alpha h+\log\left(\frac{1}{\delta}\right)\right).
\end{align}
Now we determine $\delta$ for the target total error $4\epsilon$ and total evolution time $T$.
For each $U(t_j+h,t_j)$, the circuit $\mathcal C_j$ implementing $\exp(\tilde\Omega_{(p)}(t_j+h,t_j))$ has error $ \|\mathcal C_j - U(t_j+h,t_j)\|$, hence the total error is
\begin{align}\label{eqn: circuit_error_T}
    \|\mathcal C - U(T,0) \|& 
    \leq    \|\mathcal C - e^{\tilde\Omega_{(p)}}(T,0)\| +  \| e^{\tilde\Omega_{(p)}}(T,0)- U_p(T,0)  \| +  \|U_p(T,0) - U(T,0) \|
    \\
    &\leq L \max_j \left(\|\mathcal C_j-e^{\tilde\Omega_{(p)}(t_{j+1},t_j)} \| + \|e^{\tilde\Omega_{(p)}(t_{j+1},t_j)}-e^{\Omega_{(p)}(t_{j+1},t_j)}\| \right) + \|U_p(T,0) - U(T,0) \|\notag 
    \\
    &\leq L\left(\delta+\epsilon/L + \max_j\|\tilde\Omega_{(p)}(t_j+h,t_j)-\Omega_{(p)}(t_j+h,t_j)\| \right) 
    + \epsilon
    \\ 
    &
    \leq L\delta + 3\epsilon  \tag{by \cref{lem: discretization_eqaulto_magnus_constantorder}} , 
\end{align}
 where the third inequality used \cref{lem: long_time_magnus_error_constantorder}. 
 Choose $L =\mathcal{O}{ \left( \frac{\bar \alpha_{\mathrm{comm}}^{1+1/p}\, T^{1+1/p}}{ \epsilon^{1/p}} \right)}$ as in \cref{lem: long_time_magnus_error_constantorder}, setting $L\delta=\epsilon$ by letting $\delta =\left(\frac{C_1\epsilon}{\bar \alpha_\mathrm{comm} T}\right)^{1+1/p}$ for some constant $C_1$, so the total error $\|\mathcal C-U(T,0)\|\leq 4\epsilon$.
Insert such a choice of $L$ and $\delta$ into \cref{eqn: temp_HAMT_cost_constant_p} to yield the desired HAM-T count.

By \cref{lemma: pth_order_omega}, the ancilla count for the LCU-based circuit for 
$\tilde\Omega_{(p)}(t_j+h,t_j)$ becomes 
\begin{equation}
\Or\left( p^2\log\left(\frac{pL}{\epsilon}\right)+p n_a+ p\log (M)\right).
\end{equation}
Therefore, applying QSVT+OAA~\cite{GilyenSuLowEtAl2019}, simulating the corresponding time evolution requires  
\begin{equation}
\Or\left( p^2\log\left(\frac{pL}{\epsilon}\right)+p n_a+ p\log (M)\right)
\end{equation}
ancilla qubits in total. 
Substituting  $M= \Or\left( \frac{\|A'\|\,T^{1-1/p}}{\bar \alpha_\mathrm{comm}^{1+\frac{1}{p}}\,\epsilon^{1-1/p}}\right)$ from \cref{lem: discretization_eqaulto_magnus_constantorder}, we obtain the desired scaling for the ancilla qubit count.

For the two-qubit gate counts, by~\cref{lemma: pth_order_omega} each $\tilde\Omega_{(p)}(t_j+h,t_j)$ invokes $\Or(p^3\log^2 (p)+ p\log (p) \log (M))$ number of two-qubit gates. Hence, the total number of two-qubit gate count is
\begin{align}\label{eqn: two_qubit_gate_cost_constant_p}
    L \cdot \left(C_2p^3\log (p)+C_3p\log (p)\log (M)\right) \cdot \left(C_{(p)}^{\gamma}\alpha h+\log\left(\frac{1}{\delta}\right)\right),
\end{align}
for some constants $C_2$ and $C_3$. Plugging the chosen value of $L$, and $\delta$ in \cref{eqn: two_qubit_gate_cost_constant_p} yields the desired two-qubit gate count.

The failure probability directly follows \cref{eqn: circuit_error_T}. 
\end{proof}

Next, we analyze the resource cost obtained by further optimizing the choice of the order $p$, which yields the scaling linear in $T$ and polylogarithmic in $1/\epsilon$.

\begin{cor}[Query complexity for Magnus expansion by further tuning $p$]\label{cor:optimal_query_complexity}
    Let the problem setup be the same as in \cref{thm: total_resource_cost_constantorder}, then for given $\epsilon$ and $T$, we can choose the parameters as in \cref{eqn: parameters_for_optimal_HAMT} such that there exists a quantum circuit $\mathcal C$ that implements $U_p(T,0)$ with total error $\|\mathcal C - U(T,0)\|\leq4\epsilon$, failure probability $\Or(\epsilon)$, and the following cost:
    \begin{enumerate}
    \item $\Or\left(\log^2\left(\frac{\bar\alpha_\mathrm{comm}T}{\epsilon}\right) \left(  \alpha T + \bar\alpha_\mathrm{comm}T \log\left(\frac{\bar\alpha_\mathrm{comm}T}{\epsilon}\right)\right) \right)$  uses of HAM-T oracles,
    
    \item $\Or\left(\log\left(\frac{\bar\alpha_\mathrm{comm}T}{\epsilon}\right)\cdot\left(\log\left(\frac{ \|A'\| T}{\bar{\alpha}_\mathrm{comm} \epsilon}\right)+\log\left(\frac{\bar\alpha_\mathrm{comm}T}{\epsilon}\right)\right)\right)$ ancilla qubits,
    \item $\tilde\Or\left(\left(\log^3\left(\frac{\bar\alpha_\mathrm{comm}T}{\epsilon}\right) + \log\left(\frac{\bar\alpha_\mathrm{comm}T}{\epsilon}\right)\log\left(\frac{\|A'\|T}{\bar\alpha_{\mathrm{comm}}\epsilon}\right)\right)\left(  \alpha T + \bar\alpha_\mathrm{comm}T \log\left(\frac{\bar\alpha_\mathrm{comm}T}{\epsilon}\right)\right) \right)$  uses of two-qubit gates.
\end{enumerate}
\end{cor}
\begin{proof}
    From the HAM-T cost in \cref{thm: total_resource_cost_constantorder}, we can show that it has a minimum when $p=\Or\left(\log(\frac{\bar{\alpha}_\mathrm{comm}T}{\epsilon})\right)$. In particular, one can see that
    \begin{equation}
    \left( \frac{\bar \alpha_\mathrm{comm}\,T}{\epsilon}\right)^{1/p} = \Or(1), 
    \end{equation}
    because $x^{1/\log(x)} = e$ for any positive number $x>0$.
    The results then follow directly by plugging such $p$ into the main results of \cref{thm: total_resource_cost_constantorder}.
\end{proof}

\begin{rem}
We note that in this work we do not aim to optimize the polylogarithmic factors involved here. There remains room for improvement. For example, in the QSVT + OAA lemma we applied, the dependence is kept as $\log(1/\delta)$ rather than the improved $\log(1/\delta)/\log\log(1/\delta)$; and the classical arithmetics may also be improved. Further refining the logarithmic dependence is an interesting direction for future work.
\end{rem}

\section{Conclusion}\label{sec: conclusion}

In this work, we analyzed the error scaling of the Magnus expansion at arbitrary order for simulating time-dependent quantum evolution and developed a systematic framework for implementing these algorithms on quantum circuits. 
We provided asymptotic bounds on query, qubit, and gate complexity for both constant-order and resource-optimized implementations at any given error threshold. 
To the best of our knowledge, this is the first result to establish the commutator error scaling for general time-dependent Hamiltonians.

Beyond its algorithmic implications, the Magnus series itself is an important and widely used theoretical tool for analyzing features of quantum systems as well as classical and quantum algorithms. Thus, our new analysis establishing commutator scaling may be of independent interest to a broader community.

Moreover, we propose to directly simulate the Magnus expansion as a quantum algorithm. We demonstrate its efficiency while explicitly tracking the constant dependence of the cost on the order $p$. Remarkably, we achieve a cost that scales only polynomially in $p$, despite the fact that the Magnus series involves a permutation group of size $p!$. In achieving this, we develop a state preparation method for the coefficients within Magnus expansion, a state preparation method for the time-ordered integral as well as a data lookup based SELECT oracle for Magnus expansion, which may themselves be of independent interest.

For future work, we first discuss possible theoretical improvements. 
A superconvergence phenomenon has been observed and proved for low-order Magnus algorithms~\cite{FangLiuSarkar2025,AnFangLin2022,Borns-WeilFangZhang2025}.
An interesting direction is to investigate the superconvergence properties of high-order Magnus expansion algorithms in both continuous and discrete settings. In particular, it would be valuable to examine whether certain structured Hamiltonians allow for superconvergence, which may lead to better cost scalings.
We also note that recent works~\cite{RendonWatkinsWiebe2024,Watson2024,WatsonWatkins2024} demonstrate improved cost scaling in precision through classical interpolation post-processing of observables. Such analyses typically require estimates of time derivatives in terms of the underlying time-evolved observable expectations. It would be interesting to extend our Magnus error analysis to this setting, and to evaluate the performance of our algorithm when combined with more advanced post-processing techniques for observables. 
Our current analysis addresses the worst-case scenario and applies to arbitrary initial states and observables. However, it remains open whether restricting to specific states or observables could lead to improved parameter dependence.

In practical applications with a specific problem in mind, parameters such as target precision, evolution time, and system size are fixed quantities rather than asymptotic variables. Consequently, asymptotic scaling alone may not identify the most efficient choice. In such settings, prefactors and constant overheads can play a decisive role. A natural direction for future work is to perform detailed resource estimates on concrete examples and application scenarios in order to determine the most effective algorithmic order and parameter settings.
Another open question is whether the commutator scaling can be preserved in the analysis of open quantum systems. 
Finally, it would be interesting to investigate whether this framework can also be extended to general unbounded systems.

\section*{Acknowledgments}
 D.F. and S.Z. acknowledge the support from the U.S. Department of Energy, Office of Science, Accelerated Research in Quantum Computing Centers, Quantum Utility through Advanced Computational Quantum Algorithms, grant NO. DE-SC0025572. D.F. also acknowledges the National Science Foundation CAREER Award DMS-2438074. D.L. acknowledges support from the U.S. Department of Energy (DOE) under Contract No. DE-AC02-05CH11231, through the Office of Advanced Scientific Computing Research Accelerated Research for Quantum Computing Program, MACH-Q project.
 The authors thank Bert De Jong, Zhenning Liu, Yu Tong, and Nathan Wiebe for valuable discussions.

\appendix
\section{Auxiliary estimates of Magnus error}\label{sec:appendix_pf_g_remainder}

\begin{proof}[Proof of \cref{lem:g_remainder}]
   The proof is essentially the same as \cite[Lemma 7]{FangLiuSarkar2025} and \cite[Lemma 5.1]{HochbruckLubich2003}. The underlying operator can be expressed in the Fourier space:
   \begin{align}
    g_p(\ad_{\Omega_{(p)}})(\ad_{\Omega_{(p)}}^{p+1}(\dot \Omega_{(p)}) )
    = &
    \int_\mathbb{R} \hat g_p(\xi) e^{\ad(\xi \Omega_{(p)})} \ad_{\Omega_{(p)}}^{p+1}(\dot \Omega_{(p)})  \, d\xi
    \\
    = & \int_\mathbb{R} \hat g_p(\xi) e^{\xi \Omega_{(p)}} \ad_{\Omega_{(p)}}^{p+1}(\dot \Omega_{(p)}) e^{-\xi \Omega_{(p)}} \, d\xi,
\end{align} 
where $\hat g_p \in L^1(\mathbb{R})$ is the Fourier transform of the $g_p$ satisfying
\begin{equation}
    g_p(ix) = \int_\mathbb{R} \hat g_p(\xi) e^{i\xi x} \, d\xi.
\end{equation}
Take the norm, one has
\begin{equation}
    \norm{ g_p(\ad_{\Omega_{(p)}})(\ad_{\Omega_{(p)}}^{p+1}(\dot \Omega_{(p)}) )} \leq \norm{\hat g_p}_{L^1} \norm{\ad_{\Omega_{(p)}}^{p+1}(\dot \Omega_{(p)})}.
\end{equation}
The $L^1$ norm of $\hat g_p$ satisfies
\begin{equation}
    \norm{\hat g_p}_{L^1} \leq C,
\end{equation}
for some constant $C>0$. Note that in fact it can be upper bounded by $C/(p+2)!$ for some constant $C$, but using $C$ is sufficient for our purpose.
\end{proof}

Here we present a Magnus expansion error bound in terms of infinitely many nested commutators which is easy to achieve by \cref{rem:infinite_sum}. Note that our main result of the Magnus error (\cref{thm:magnus_trunction_finite_comm}) is different as it only depends on a finite number of nested commutators.
\begin{thm}[Alternative Error Bound of $p$-th Order Magnus Expansion] \label{thm:magnus_trunction_infty_comm}
The local truncation error of the $p$-th order Magnus expansion $e_\mathrm{loc}(h)$ on the short-time interval $[t_j, t_j+h]$ satisfies
\begin{equation}
    e_\mathrm{loc}(h) = \norm{U(t_j+h, t_j) - U_p(t_j+h, t_j)} \leq \sum_{n = p+1}^\infty \frac{1}{n^2}\alpha_{\mathrm{comm},n} h^n \leq 2 \tilde \alpha_{\mathrm{comm}}^{p+1}h^{p+1},
\end{equation}
for $h$ such that $h \tilde\alpha_\mathrm{comm} < 1$,
where 
\begin{equation}
    \tilde \alpha_{\mathrm{comm}}: = \max_{q \geq p+1} \{\alpha_{\mathrm{comm},q}^{1/q}\},
\end{equation}
with $\alpha_{\mathrm{comm},q}$ given as \cref{eq:def_alpha_comm_q}.
In this case, the global error is bounded by 
\begin{equation}
     \norm{U(T,0) - U_p(T,0)} \leq   2 \tilde \alpha_{\mathrm{comm}}^{p+1} h^{p}T.
\end{equation}

\end{thm}

\bibliographystyle{unsrt}
\bibliography{magnus-2.bib}
\end{document}